\documentclass[aps,twocolumn,pre,floatfix]{revtex4-2}

\usepackage{xcolor,hyperref}
\hypersetup{
   colorlinks,
   linkcolor={blue!50!black},
   citecolor={blue!50!black},
   urlcolor={blue!80!black}
}

\usepackage{epsfig, amsmath}
\usepackage{bm}
\usepackage{soul,ulem}
\usepackage{hyperref}
\usepackage{footmisc}
\usepackage{wrapfig}
\usepackage{amssymb}
\DeclareMathAlphabet{\mathitbf}{OML}{cmm}{b}{it}

\newcommand{\ket}[1]{|#1\rangle}
\newcommand{\bra}[1]{\langle #1|}
\newcommand{\braket}[2]{\langle #1|#2\rangle}

\newcommand{\dv}{\mathitbf d}

\newcommand{\xv}{\mathitbf x}

\newcommand{\nv}{\mathitbf n}
\newcommand{\Dv}{\mathitbf D}

\newcommand{\calBold}[1]{\mbox{\boldmath${\cal #1}$}}
\newcommand{\dbar}{{\,\mathchar'26\mkern-12mu d}}

\definecolor{darkGreen}{RGB}{0,100,0}

\begin{document}

\title{Elasticity of self-organized frustrated disordered spring networks}
\author{Tommaso Pettinari$^{1,2}$}
\email{t.pettinari@uva.nl}
\author{Gustavo During$^{3}$}
\email{gduring@fis.puc.cl}
\author{Edan Lerner$^{1}$}
\email{Corresponding author: e.lerner@uva.nl}
\affiliation{$^{1}$Institute of Theoretical Physics, University of Amsterdam, Science Park 904, 1098 XH Amsterdam, the Netherlands\\
$^{2}$Van der Waals-Zeeman Institute, University of Amsterdam,
Science Park 904, 1098 XH Amsterdam, the Netherlands\\
$^{3}$Instituto de F\'isica, Pontificia Universidad Cat\'olica de Chile, 8331150 Santiago, Chile}

\begin{abstract}
There have been some interesting recent advances in understanding the notion of mechanical disorder in structural glasses and the statistical mechanics of these systems' low-energy excitations. Here we contribute to these advances by studying a minimal model for structural glasses' elasticity in which the degree of mechanical disorder --- as characterized by recently introduced dimensionless quantifiers --- is readily tunable over a very large range. We comprehensively investigate a number of scaling laws observed for various macro-, meso- and microscopic elastic properties, and rationalize them using scaling arguments. Interestingly, we demonstrate that the model features the universal quartic glassy vibrational density of states as seen in many atomistic and molecular models of structural glasses formed by cooling a melt. The emergence of this universal glassy spectrum highlights the role of self-organization (towards mechanical equilibrium) in its formation, and elucidates why models featuring structural frustration alone do not feature the same universal glassy spectrum. Finally, we discuss relations to existing work in the context of strain-stiffening of elastic networks and of low-energy excitations in structural glasses, in addition to future research directions.


\end{abstract}

\maketitle

\section{Introduction}

Many static and dynamic mechanical properties of structural glasses depend upon their featured degree of mechanical disorder. Some well-known examples manifesting this dependence include wave attenuation rates~\cite{Schirmacher_prl_2007,Schirmacher_2013_boson_peak,scattering_jcp,grzegorz_soft_matter_2020,jcp_letter_scattering_2021,massimo_scipost_2023}, elasto-plastic responses~\cite{falk_shi_prl_2005,Bouchbinder2009c,Ozawa6656,francescos_and_yoshino_science_advances_2018} and fracture mechanics~\cite{Rycroft2012,Eran_mechanical_glass_transition,david_fracture_mrs_2021}. Consequently, many efforts attempting to identify useful and broadly applicable quantifiers of mechanical disorder in structural glasses were put forward in recent years~\cite{falk_prl_2016,david_collaboration_2020,chi_paper_2023,ml_roadmap_2023,david_pairwise_stuff_2023}. These efforts complement a different class of approaches --- applied predominantly to understanding the structure-dynamics relations in supercooled liquids (see e.g.~\cite{paddy_huge_review_2015,Tong2019}) --- in which glassy disorder is quantified via positional disorder (of the relevant constituents) alone, with little or no reference to mechanics. The role of disorder in the physics of glassy solids has been incorporated in various theoretical frameworks e.g.~the Shear Transformation Zones theory~\cite{Bouchbinder2009b} and Heterogeneous Elasticity Theory~\cite{Schirmacher_prl_2007,Schirmacher_2013_boson_peak}. 

One key challenge towards a thorough understanding of the elasticity of structural glasses involves identifying simple models or glass-formation protocols (or both) that allow workers to tune the degree of mechanical disorder of the model systems over a large range. This has been previously accomplished in Refs.~\cite{swap_prx_MW,fsp} by allowing for additional degrees of freedom in the Hamiltonian of a generic glass forming model --- in this case, the particles' effective sizes ---, and subsequently freezing those additional degrees of freedom after glass formation. Computer glasses formed with this protocol showed remarkable stress-strain curves with very pronounced stress-overshoots, which are reminiscent of those seen upon shear deformation of ultra-stable glasses~\cite{Ozawa6656,francescos_and_yoshino_science_advances_2018}. They also feature an enormous variability of the width-to-mean ratio of the sample-to-sample distribution of the shear modulus --- denoted there, here and in what follows as $\chi$ ---, shown to vary by over a factor of 4 by tuning the stiffness associated with the additional degrees of freedom. This approach to tuning mechanical disorder in computer glasses was also employed in Ref.~\cite{david_fracture_mrs_2021} where the said variability gives rise to a ductile-to-brittle transition in glassy samples under uniaxial tension. 

\begin{figure*}[ht!]
 \includegraphics[width = 0.98\textwidth]{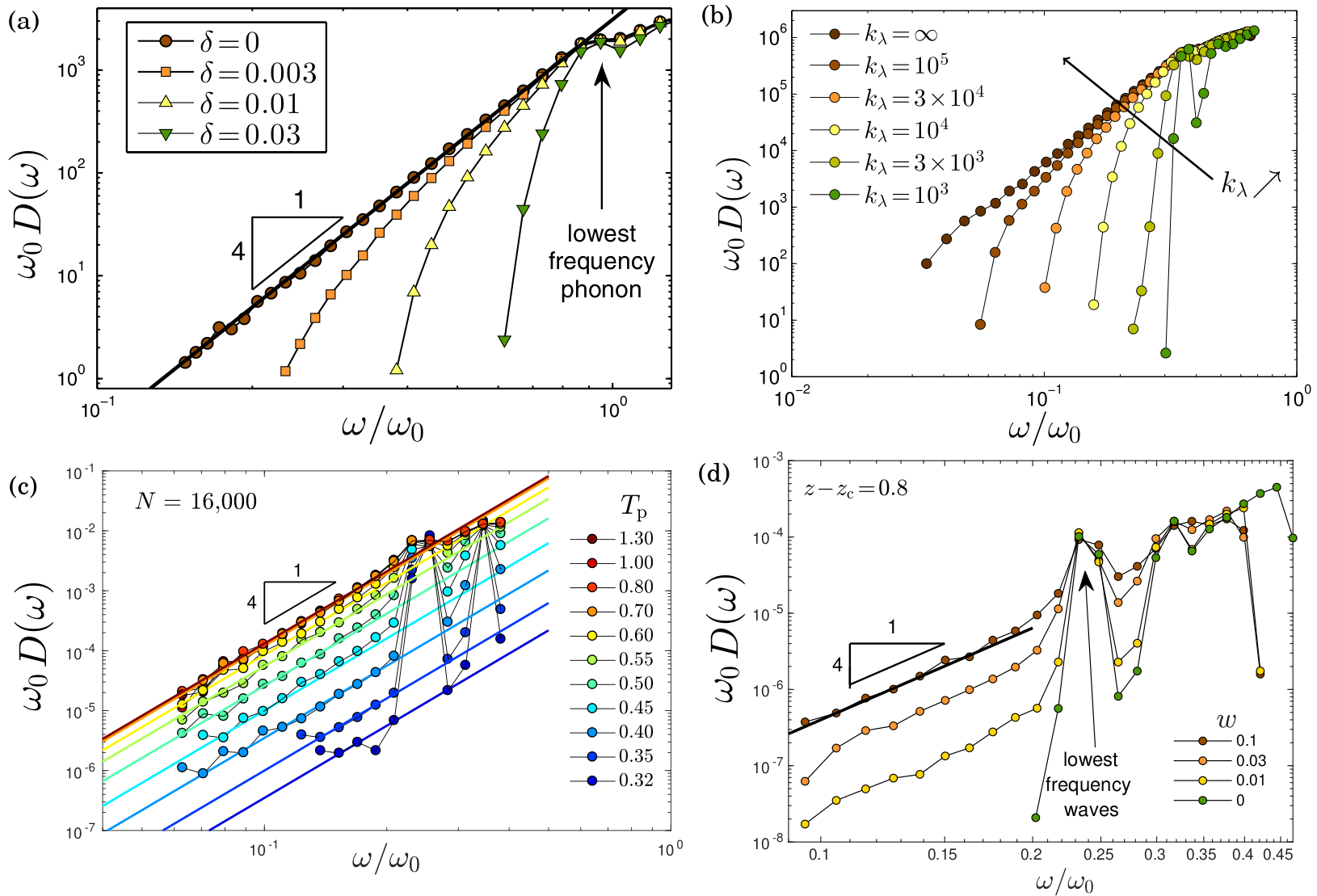}
 \vspace{-0.4cm}
  \caption{\footnotesize (a) The vibrational density of states (VDoS) calculated for computer glasses in which the interparticle forces were factored by $1\!-\!\delta$, see legend for values of $\delta$. Setting $\delta\!>\!0$ results in a gap $\sim\!\sqrt{\delta}$ in the VDoS~\cite{inst_note}. Here and in (b)-(d), $\omega_0\!\equiv\!c_{\rm s}/a_0$ where $c_{\rm s}$ is the speed of shear waves and $a_0$ an interparticle length. Adopted from Ref.~\cite{inst_note}. (b) VDoS of computer glasses in which the particles' effective sizes are allowed to vary under a potential of characteristic stiffness $k_\lambda$ \emph{only} during glass formation (and frozen thereafter). A gap $\sim\!1/\sqrt{k_\lambda}$ opens in the VDoS. Adopted from Ref.~\cite{fsp}. These two examples demonstrate that some approaches to tuning mechanical disorder in computer glasses result in anomalous vibrational spectra compared to those of glasses quenched from a melt -- that universally feature $\sim\!\omega^4$ nonphononic spectra, as shown in panel (c) for glasses made with the Swap-Monte-Carlo method~\cite{LB_swap_prx}. Adopted from Ref.~\cite{phonon_widths2}. In this work we present a model and sample-formation protocol that both allows for a large tunability of mechanical disorder, \emph{and} features the universal quartic nonphononic spectrum as shown in panel (d) and discussed in Sect.~\ref{sec:summary} below.}
  \label{fig:other_approaches}
\end{figure*}

Another route towards tuning mechanical disorder in simple models was put forward in Refs.~\cite{eric_boson_peak_emt,breakdown,inst_note,jcp_letter_scattering_2021,chi_paper_2023}; there, the internal stresses of a conventional computer glass were artificially reduced, leading to smaller mechanical fluctuations in glassy samples. Refs.~\cite{jcp_letter_scattering_2021,chi_paper_2023} showed that this reduction of internal stresses can give rise to almost an order of magnitude variability of the mechanical-disorder quantifier $\chi$. Using this approach, in Ref.~\cite{jcp_letter_scattering_2021} it was shown that  attenuation rates of shear waves scales as $\chi^2$, in agreement with the predictions of Heterogeneous Elasticity Theory~\cite{Schirmacher_prl_2007,Schirmacher_2013_boson_peak}.

While the aforementioned  exercises and protocols are undoubtedly useful and insightful, they produce glassy structures that cannot represent well the class of configurations obtained by cooling a liquid into a glass. In particular, glasses formed by quenching a liquid have been shown to feature a universal gapless nonphononic spectra ${\cal D}(\omega)\!\sim\!\omega^4$~\cite{soft_potential_model_01,soft_potential_model_02,JCP_Perspective} (here $\omega$ is an angular frequency), independent of glass formation history~\cite{LB_modes_2019,pinching_pnas,footnote_finite_size_effects}, spatial dimension~\cite{modes_prl_2018}, or any microscopic details~\cite{modes_prl_2020}. In contrast, glassy structures formed with the aforementioned protocols --- either by adding degrees of freedom during glass formation~\cite{swap_prx_MW,fsp}, or by artificially reducing the internal stresses~\cite{eric_boson_peak_emt,breakdown,inst_note,jcp_letter_scattering_2021,chi_paper_2023} --- feature a \emph{gapped} nonphononic spectra (see examples in Fig.~\ref{fig:other_approaches}), rendering them less realistic and thus less relevant to natural or laboratory glasses. 

In this work we study a minimal off-lattice model system and glass-formation protocol that --- when employed together --- allow for a very large tunability of mechanical disorder, while also featuring the universal $\sim\!\omega^4$ nonphononic spectrum as seen in computer glasses formed by quenching a melt, see Fig.~\ref{fig:other_approaches} and associated caption. We study various micro-, meso- and macroscopic observables as a function of the model's key parameters, and rationalize the observed scaling laws using scaling ansatze and  arguments. We further discuss the relation of our work to a recently introduced class of mean-field models for glassy excitations~\cite{scipost_mean_field_qles_2021,meanfield_qle_pierfrancesco_prb_2021}, to the phenomenon of strain-stiffening of elastic networks~\cite{robbie_nature_physics_2016,robbie_pre_2018,strain_stiffening_2023}, and to other recent relevant work. Finally, future research directions are proposed.

\begin{figure*}[ht!]
\includegraphics[width = 0.85\textwidth]{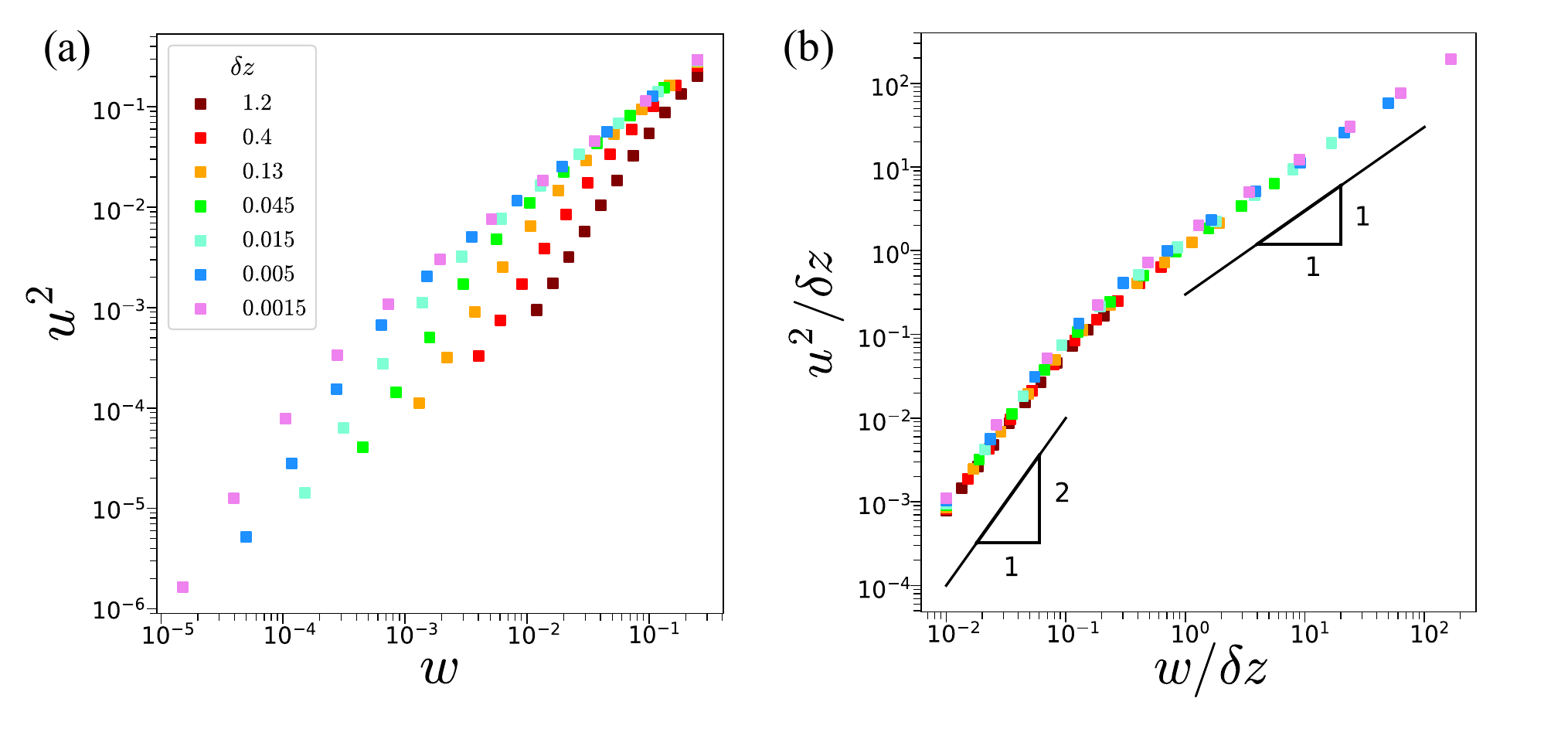}
  \caption{\footnotesize Mean squared displacements $u^2$ between the relaxed, initial network, and the network obtain after introducing the mechanical frustration and minimizing the energy, as described in the text. Panel (a) shows the raw data for various coordinations $z$ and widths $w$ (see text for definitions). Panel (b) employs the scaling form of Eq.~(\ref{eq:displacements}) to find a convincing data collapse, validating the predicted crossover between ballistic and diffusive behavior. This form indicates that, at high $w$'s, observables becomes $z$ independent, see text for discussion.}
  \label{fig:displacements}
\end{figure*}

\section{Model and sample-formation protocol}

We consider a minimal model consisting of a disordered Hookean-spring networks in three dimensions (3D), connected at each of its $N$ nodes to point-like unit masses. The networks analyzed are constructed by adopting the contact network between particles of well-compressed packings of soft (harmonic) spheres. The edges of the adopted network are pruned to reach some target mean coordination $z$; in order to avoid dangling nodes or other unusual structural fluctuations, the pruning is done following the scheme described in Ref.~\cite{anomalous_elasticity_soft_matter_2023} that reduces node-to-node coordination fluctuations. In this edge-pruned initial spring network, all of the springs' restlengths are set to be equal to those springs' actual lengths, and so there is initially no energy, stresses nor frustration in the network. The springs all share the same stiffness $k$, set to unity in what follows. We use $a_0\!\equiv\!(V/N)^{1/\dbar}$ as the units of length (set to unity), such that, in what follows, energies are expressed in terms of $ka_0^2$, and stresses, pressures and elastic moduli in terms of $k/a_0$.

In the next step of constructing glassy samples from our simple model, we introduce shifts $\delta\ell_{ij}$ in the restlengths $\ell_{ij}$ of the springs. The restlength-shifts $\delta\ell_{ij}$ are drawn from a Gaussian distribution with zero mean and width $w$, the latter forms one of the two key parameters of this model system (along with the coordination $z$). The introduction of shifts in the springs' rest-lengths leads to mechanical imbalance, which is then eliminated by a potential-energy minimization~\cite{fire}, during which the system self-organizes into a state satisfying mechanical equilibrium. To avoid finite-size effects, we used the scheme described above to construct mechanically frustrated networks with $N\!=\!62500$ nodes; unless stated otherwise, the data below is reported for this system size.

The model and glass-formation protocol described above are identical to one of the three variants put forward very recently in~\cite{liu_arxiv_frustrated_networks_2023}; in that work the focus is on the floppy regime $z\!<\!z_{\rm c}\!=\!2\dbar$ with $\dbar$ denoting the dimension of space. Our model and protocol also bear similarity to the protocol and glass former employed in Ref.~\cite{massimo_scipost_2023} to create glassy configurations over a wide range of mechanical disorder, which also features the universal $\sim\!\omega^4$ nonphononic spectrum. We discuss our results in the context of these prior works in the discussion Section.

\section{Results}
\label{sec:results}

\subsection{Notation and formalism}
\label{sec:formalism}

Before presenting our numerical results, we briefly review some of the formalism used to rationalize our observations. In particular, we adopt the bra-ket notation of Refs.~\cite{asm_pnas_2012,sss_epje_2018} and make use of the equilibrium matrix ${\cal S}^T$~\cite{calladine_buckminster_1978} that takes the vector sum of pairwise (spring) forces $\ket{f}$ exerted on each node, and results in a vector $\ket{F}$ of the \emph{net} force on those nodes, namely
\begin{equation}
\ket{F} = {\cal S}^T\ket{f}\,.
\end{equation}

Furthermore, we make use of the known form of the nonphononic vibrational density of states ${\cal D}(\omega)$ in \emph{relaxed} spring networks at some coordination $z$, namely that a plateau of extended anomalous vibrational modes emerges from a frequency $\omega_\star\!\sim\!z\!-\!z_{\rm c}\!\equiv\!\delta z$~\cite{matthieu_thesis,phonon_gap_2012,Beltukov2015,silbert_pre_2016}.

Finally, denoting the potential energy by $U(\xv)$ and coordinates by $\xv$, we use that the Hessian matrix ${\cal H}\!\equiv\!\frac{\partial^2 U}{\partial\xv\partial\xv}$ of a relaxed spring network (of springs of unit stiffness, as in our model) reads~\cite{asm_pnas_2012}
\begin{equation}
    {\cal H} = {\cal S}^T{\cal S}\,.
\end{equation}
Furthermore, the eigenvectors $\ket{\Psi_l}$ of ${\cal H}$ satisfy~\cite{asm_pnas_2012} \begin{equation}\label{eq:es_on_eigenvector}
{\cal S}\ket{\Psi_l}=\omega_l\ket{\varphi_l}\,,
\end{equation}
where $\ket{\varphi_l}$ is an eigenvector of the matrix ${\cal S}{\cal S}^T$ associated with the same eigenvalue $\omega_l^2$.

\begin{figure*}[ht!]
\includegraphics[width = 0.8\textwidth]{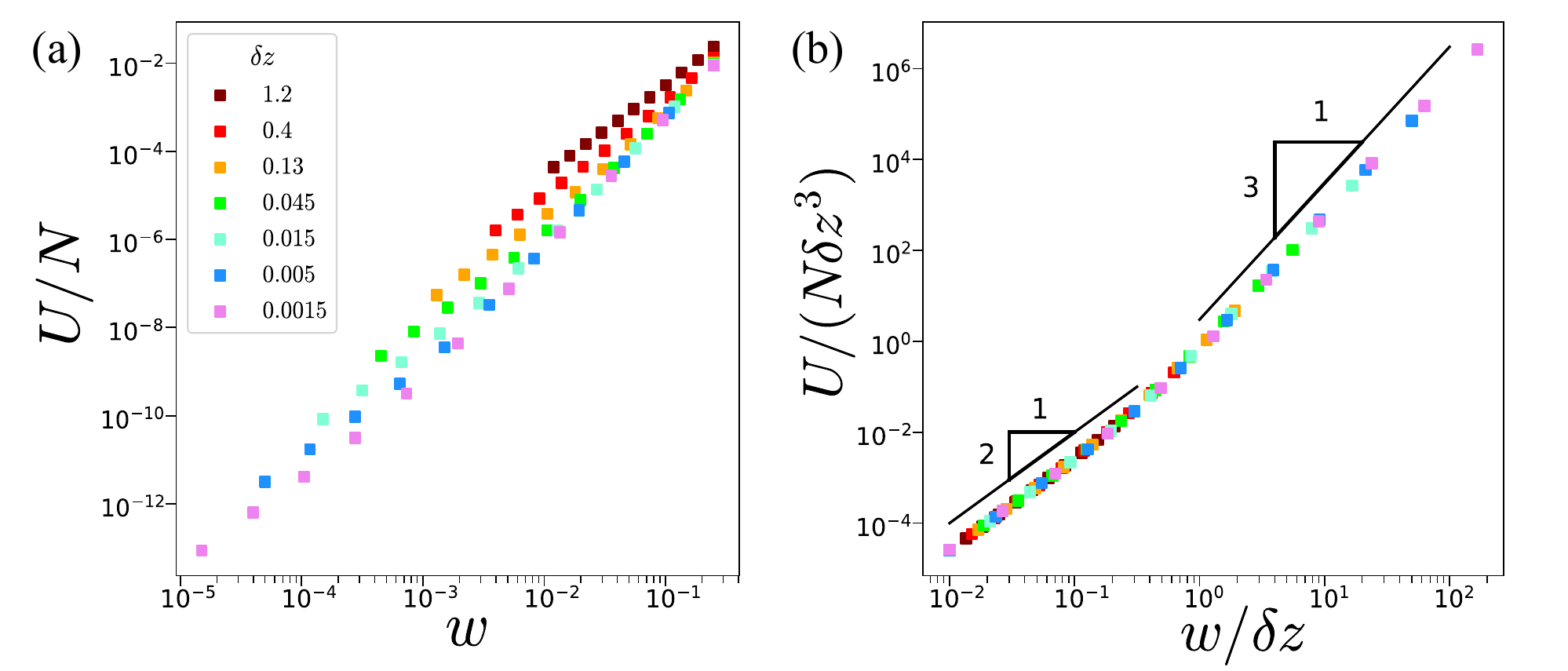}
  \caption{\footnotesize (a) Potential energy per particle $U/N$ plotted under variations of the width $w$ and various coordinations $z$ as detailed in the legend. (b) The scaling form Eq.~(\ref{eq:potential_scaling_form}) leads to a convincing collapse, see text for further details.}
  \label{fig:potential}
\end{figure*}

\subsection{Displacements towards mechanical equilibrium}
\label{sec:displacements}

The first observable we consider is the mean-squared displacements $u^2$ between the initial network --- before the restlength shifts are introduced --- and the force-balanced configurations obtained after introducing the restlength shifts and minimizing the energy. Assuming the springs all share the same unit stiffnesses, the \emph{net} force $\ket{F}$ on the nodes due to introducing the restlength shifts $\ket{\delta \ell}$ can be written as
\begin{equation}
\ket{F} = {\cal S}^T\ket{\delta \ell} = w\,{\cal S}^T\ket{\eta}\,,
\end{equation}
where we have defined $\ket{\delta\ell}\!=\!w\ket{\eta}$ and $\eta$ is a vector of $Nz/2$ uncorrelated, normally distributed random variables with unit variance and zero mean.

The linear displacement response to introducing the restlength shifts is therefore
\begin{equation}
    \ket{u} = {\cal H}^{-1}\ket{F} = w{\cal H}^{-1}{\cal S}^T\ket{\eta}\,,
\end{equation}
and therefore its square magnitude follows~\cite{breakdown}
\begin{eqnarray}
    u^2 & = & \braket{u}{u} = \bra{F}{\cal H}^{-2}\ket{F} = w^2\bra{\eta}{\cal S}{\cal H}^{-2}{\cal S}^T\ket{\eta} \nonumber \\
    & = &\! w^2\! \sum_l \frac{\bra{\eta}{\cal S}\ket{\Psi_l}\bra{\Psi_l}{\cal S}^T\ket{\eta}}{\omega_l^4} = w^2 \sum_l \frac{\bra{\eta}{\cal S}\ket{\Psi_l}^2}{\omega_l^4} \nonumber \\
    & = & \!w^2\! \sum_l\! \frac{\braket{\eta}{\varphi_l}^2}{\omega_l^2}\!\sim\! w^2 \!\!\!\int\limits_{\delta z}\!\!\frac{D(\omega)d\omega}{\omega^2}\!\sim\! \frac{w^2}{\delta z}\sim \delta z\! \left(\frac{w}{\delta z}\right)^2\!\!\!,
\end{eqnarray}
where we have suppressed the trivial $\sim\!N$ dependence, and used Eq.~(\ref{eq:es_on_eigenvector}). The above argument implies that, in the linear regime, $u^2/\delta z$ should be a quadratic function of the ratio $w/\delta z$, and that the relevant scale for the restlength shifts in our system is $w_\star\!\sim\!\delta z$. Indeed, in the linear-response regime, we expect a ballistic behavior of displacements (where the role of time is played by $w$), consistent with the above arguments.

Given the above result, we make the scaling ansatz
\begin{equation}\label{eq:displacements}
u^2 \sim \delta z\, {\cal F}_u(w/\delta z)\,,
\end{equation}
where the scaling function ${\cal F}_u(y)$ satisfies ${\cal F}_u(y)\!\sim\!y^2$ for $y\!\equiv\!w/\delta z\!\ll\!1$. Furthermore, since one expects that the nonlinear displacement response should become \emph{diffusive}, we expect ${\cal F}_u(y)\!\sim\!y$ for $y\!\gg\!1$.

All the above predictions are validated in Fig.~\ref{fig:displacements}; we find that the scaling form Eq.~(\ref{eq:displacements}) leads to a convincing data-collapse. Importantly, we further note that, in the nonlinear (diffusive) regime $w\!\gg\!\delta z$, Eq.~(\ref{eq:displacements}) indicates that $u^2\!\sim\!w$, \emph{independent of the coordination $z$}. We therefore postulate that, since displacements become coordination-independent in the nonlinear regime, then other observables of interest will also show $z$-independence for $w\!\gg\!\delta z$. In what follows, we will use this assumption of $z$-independence, together with linear response arguments, to obtain the scaling exponents of various observables in the nonlinear, $w\!\gg\!\delta z$ regime.

\subsection{Potential energy}

The potential energy $U$ of our model reads
\begin{equation}
U(w) = \sum_{\mbox{\tiny springs $i,j$}}(r_{ij}-\ell_{ij}(w))^2\,,
\end{equation}
where the restlenghs $\ell_{ij}(w)\!=\!\ell_{ij}^{(0)}\!+\!\delta\ell_{ij}(w)$ with $\delta\ell_{ij}(w)\!=\!w\eta_{ij}$ and $\eta_{ij}$ are normally distributed, uncorrelated random variables with unit variance and zero mean. 
In the harmonic regime, the potential energy is obtained by approximating $    U(w)\!\simeq\!\frac{d^2 U}{dw^2}\big|_{w=0}w^2$,
where the total derivative $d/dw$ are taken under the preservation of mechanical equilibrium~\cite{lutsko}. In addition, we recall that since $U(w\!=\!0)\!=\!0$, together with that the reflection symmetry $w\!\to\!-w$ implies that $dU/dw\!=\!0$, one deduces the quadratic scaling of the energy with $w$.

\begin{figure*}[ht!]
 \includegraphics[width = 0.8\textwidth]{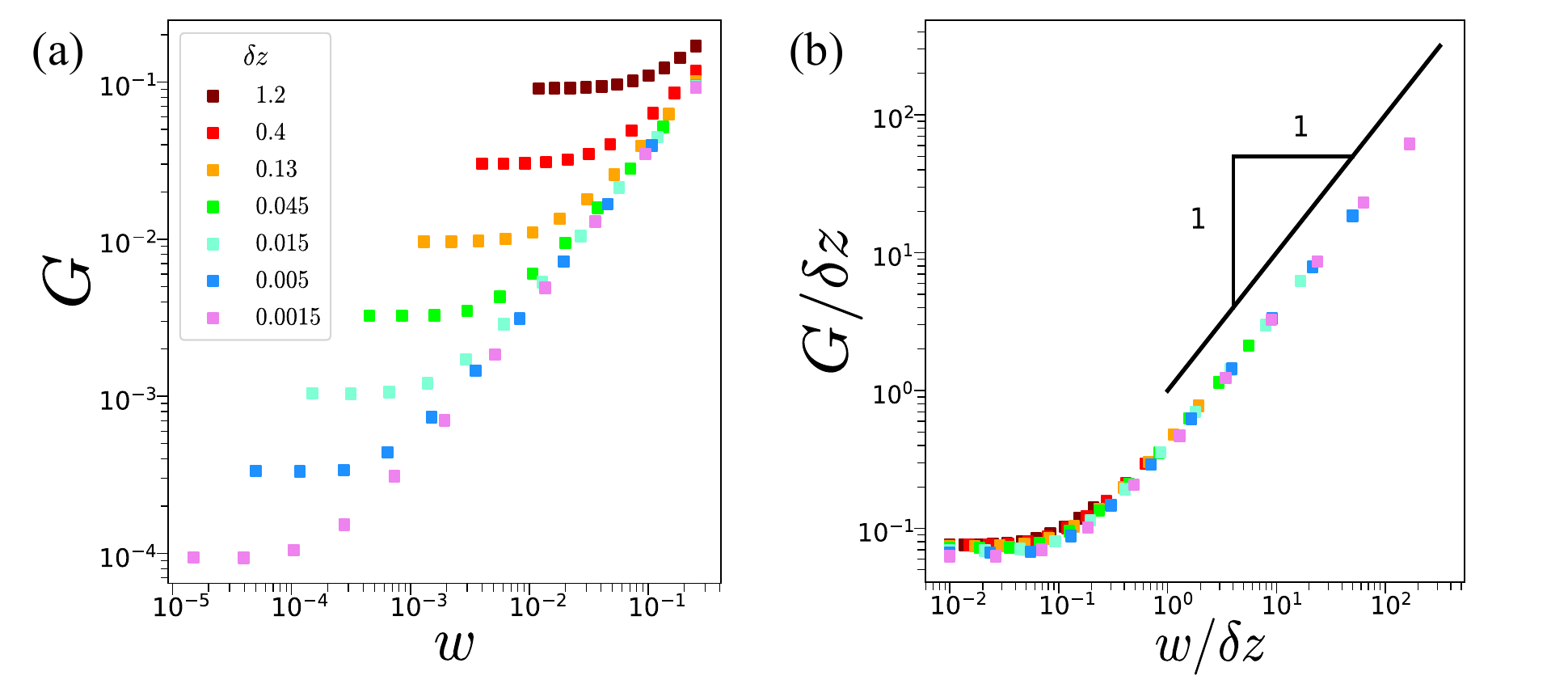}
  \caption{\footnotesize (a) Shear modulus $G$ plotted under variations of the width $w$ and various coordinations $z$ as detailed in the
legend. (b) The scaling form Eq.~(\ref{eq:shear_modulus_scaling_form}) leads to a convincing collapse, see text for further details.}
  \label{fig:shear_modulus}
\end{figure*}

The second derivative reads~\cite{matthieu_thesis,Ellenbroek_2009,sss_epje_2018}
\begin{eqnarray}
    \frac{d^2 U}{dw^2}\bigg|_{w=0}\!\!\!\!\!\! & = & \braket{\eta}{\eta}-\bra{\eta}{\cal S}{\cal H}^{-1}{\cal S}^T\ket{\eta}\nonumber \\
    & = & \bra{\eta}\big({\cal I} - {\cal S}{\cal H}^{-1}{\cal S}^T\big)\ket{\eta}\sim N\delta z\,,\nonumber 
\end{eqnarray}
and so we expect that, for small $w$
\begin{equation}
    \frac{U(w)}{N} \sim  \delta z\, w^2 \sim  \delta z^3  \!\left(\frac{w}{\delta z}\right)^2\,.
\end{equation}
The linear response result derived above suggests that the energy per particle is given by the following scaling form
\begin{equation}\label{eq:potential_scaling_form}
    U/N = \delta z^3{\cal F}_U(w/\delta z)\,,
\end{equation}
where ${\cal F}_U(y)\!\sim\!y^2$ for $y\!\ll\!1$. As discussed above, if we accept the postulation that observables must become independent of $z$ at large $w$, then we expect ${\cal F}_U(y)\!\sim\!y^3$ for $y\!\gg\!1$, or  $U/N\!\sim\!w^3$ for $w\!\gg\!\delta z$. These predictions are validated in Fig.~\ref{fig:potential}.

\subsection{Shear modulus}
\label{sec:shear_modulus}

We next analyze the shear modulus $G$ of our glassy configurations; microscopic expressions are available e.g.~in~\cite{lutsko}. In the unjamming literature it is well-known that, for relaxed spring networks at coordination $z$, $G\!\sim\!\delta z$. We thus make the following scaling ansatz 
\begin{equation}\label{eq:shear_modulus_scaling_form}
    G\sim\delta z\, {\cal F}_G(w/\delta z)\,, 
\end{equation}
where ${\cal F}_G(y)\!\to\!\mbox{const}$ for $y\!\ll\!1$. Demanding once again that $G$ becomes $z$-independent at large $w$, we predict $G\!\sim\!w$ at for $w\!\gg\!\delta z$, and thus ${\cal F}_{G}(y)\!\sim\!y$ for $y\!\gg\!1$. These predictions are validated in Fig.~\ref{fig:shear_modulus}. 

\subsection{Characteristic mesoscopic elastic stiffness}
\label{sec:meso_stiffness}

After having analyzed the macroscopic shear modulus $G$ in the previous subsection, it is interesting to consider how the mesoscopic stiffness associated with \emph{local} perturbations behaves in our model systems. Interestingly, in Refs.~\cite{cge_paper,pinching_pnas,sticky_spheres_part2} it was shown that mesoscopic stiffnesses of the type discussed in what follows are more susceptible to thermal annealing compared to macroscopic elastic moduli, in simple models of computer glasses quenched from a melt. Similar ideas and their implications on supercooled liquids' vibrational entropy and fragility were put forward earlier in Ref.~\cite{wyart_vibrational_entropy}. Finally, in the unjamming literature it has been shown~\cite{matthieu_PRE_2005,matthieu_thesis,mw_EMT_epl,new_variational_argument_epl_2016} that the stiffness $\omega_\star^2\!\sim\!\delta z^2$ associated with anomalous vibrational modes in disordered, relaxed spring networks vanishes more quickly upon approaching the unjamming point, compared to the macroscopic shear modulus -- which vanishes as $\delta z$. These previous results indicate that mesoscopic and macroscopic stiffnesses quite generically feature different dependencies on glass control parameters, in several scenarios.  

\begin{figure*}[ht!]
 \includegraphics[width = 0.75\textwidth]{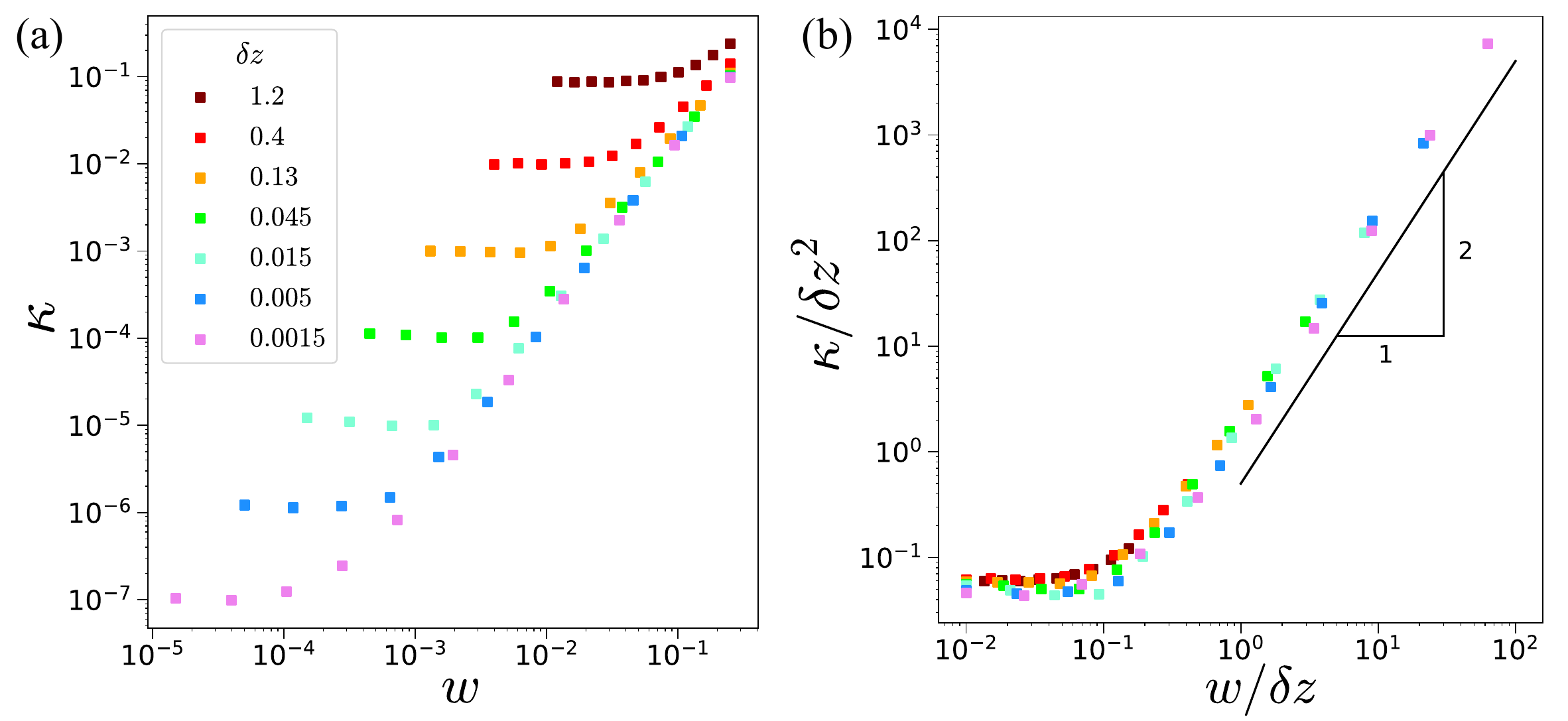}
  \caption{\footnotesize (a) Mesoscopic elastic stiffness $\kappa$ plotted under variations of the amplitude of the restlength noise $w$ and various coordinations $z$ as detailed in the legend. (b) The
scaling form Eq.~(\ref{eq:kappa_scaling}) leads to a convincing collapse, see text for further details.}
  \label{fig:kappa}
\end{figure*}

In order to probe the mesoscopic stiffness in our model, we choose a random pair of nodes $i,j$ that are positioned close to each other, but are \emph{not} connected by a spring. We apply a unit force dipole of the form $\dv_k^{ij}\!=\!(\delta_{jk}\!-\!\delta_{ik})\nv_{ij}$ to the selected pair of nodes (here $\nv_{ij}$ is the unit vector pointing from node $i$ to node $j$), and obtain the displacement response $\Dv_k^{ij}\!=\!\calBold{H}^{-1}_{k\ell}\!\cdot\!\dv_\ell^{ij}$ to the force dipole. The stiffness $\kappa_{ij}$ associated with this perturbation is then
\begin{equation}\label{eq:kappa}
    \kappa_{ij} = \frac{\Dv^{ij}\cdot\calBold{H}\cdot\Dv^{ij}}{\Dv^{ij}\cdot\Dv^{ij}} =  \frac{\dv^{ij}\cdot\calBold{H}^{-1}\cdot\dv^{ij}}{\dv^{ij}\cdot\calBold{H}^{-2}\cdot\dv^{ij}}\,.
\end{equation}
In what follows we consider the average $\kappa\!\equiv\!\langle \kappa_{ij}\rangle_{ij}$ over a random selection of pairs $i,j$ in our ensemble of glassy samples. According to Ref.~\cite{sss_epje_2018}, in relaxed spring networks (i.e.~for $w\!=\!0$ in our model) the numerator of Eq.~(\ref{eq:kappa}) is expected to scale as $\delta z$, and according to Refs.~\cite{breakdown,new_variational_argument_epl_2016} the denominator of Eq.~(\ref{eq:kappa}) should scale as $1/\delta z$, leaving us with the prediction $\kappa\!\sim\!\delta z^2$ for $w\!\to\!0$, in agreement with the arguments of Ref.~\cite{new_variational_argument_epl_2016} for $\omega_\star^2\!\sim\!\delta z^2$.

What happens to the mesoscopic stiffness $\kappa$ upon introducing mechanical frustration into the network? As done above, we make the scaling ansatz
\begin{equation}\label{eq:kappa_scaling}
\kappa \sim \delta z^2 \,{\cal F}_\kappa(w/\delta z)\,,
\end{equation}
such that ${\cal F}_\kappa(y)\!\to\!\mbox{const}$ for $y\!\ll\!1$. Requiring that $\kappa$ becomes $z$-independent at large $w$, we predict that ${\cal F}_\kappa(y)\!\sim\!y^2$ and so we expect that $\kappa\!\sim\!w^2$ at large $w$. This result should be compared (in the context of the discussion above) with $G\!\sim\!w$ in the same, nonlinear regime, indicating that the relation $\kappa\!\sim\!G^2$ holds both in the linear and nonlinear regimes. A similar universality was recently pointed out in Ref.~\cite{bending_networks_arXiv_2023} for spring networks endowed with weak bending interactions. Our predictions are verified in Fig.~\ref{fig:kappa}.

\subsection{Mechanical disorder}

\begin{figure*}[ht!]
 \includegraphics[width = 0.8\textwidth]{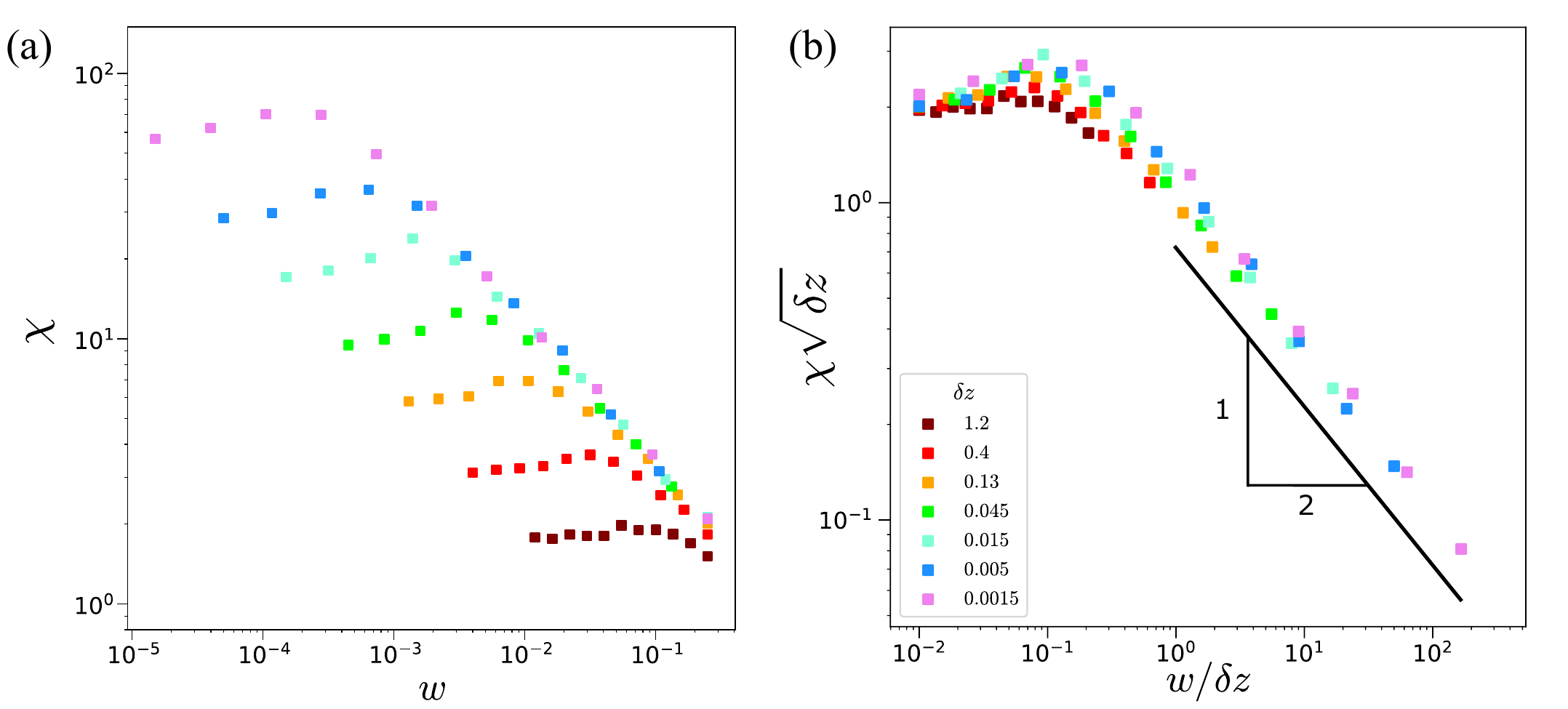}
  \caption{\footnotesize (a) Dimensionless mechanical disorder quantifier $\chi$ plotted under variations of the amplitude of the restlength noise $w$ and various coordinations $z$ as detailed in the legend. (b) The scaling form Eq.~(\ref{eq:chi_scaling}) leads to a convincing collapse, see text for further details.}
  \label{fig:chi}
\end{figure*}

We end this Section with the behavior of the dimensionless quantifier $\chi$ of mechanical disorder, defined as~\cite{phonon_widths,phonon_widths2,jcp_letter_scattering_2021,chi_paper_2023,julia_chi_jcp}
\begin{equation}
    \chi \equiv \sqrt{N}\frac{\mbox{std(G)}}{\mbox{mean}(G)}\,,
\end{equation}
where $\mbox{std}(\bullet)$ stands for the \emph{ensemble} standard deviation, and $\mbox{mean}(\bullet)$ refers to the ensemble mean. The $\sqrt{N}$ factor ensures $\chi$ is $N$-independent (for large-enough $N$). In previous work it has been shown that for relaxed spring networks with coordination $z$, $\chi\!\sim\!1/\sqrt{\delta z}$~\cite{chi_paper_2023,julia_chi_jcp}. We therefore make the scaling ansatz that
\begin{equation}\label{eq:chi_scaling}
\chi\sim\delta z^{-1/2}\,{\cal F}_\chi(w/\delta z)\,,
\end{equation}
where ${\cal F}_\chi(y)\!\to\!\mbox{const}$ for small $y$. Requiring that $\chi$ becomes $z$-independent at large $w$ implies $\chi\!\sim\!1/\sqrt{w}$ at large $w$, and so ${\cal F}_\chi(y)\!\sim\!y^{-1/2}$ for $y\!\gg\!1$. All these predictions are validated in Fig.~\ref{fig:chi}. We note that, for estimating $\mbox{std}(G)$ we employed that outlier elimination method as explained in Ref.~\cite{phonon_width_2}.

\section{Summary, Discussion and outlook}
\label{sec:summary}

In this work we have studied and rationalized the elastic properties of a minimal model for the elasticity of glassy solids in which the degree of mechanical disorder can be tuned over a large range, cf.~Fig.~\ref{fig:chi}a. At the same time, and different from some other models for glass elasticity, this minimal model features the universal $\sim\!\omega^4$ nonphononic spectrum (cf.~Fig.~\ref{fig:other_approaches}), which implies that it can better represent more realistic glassy solids compared to other computational models in which mechanical disorder is tunable~\cite{eric_boson_peak_emt,breakdown,inst_note,jcp_letter_scattering_2021,chi_paper_2023}. Our approach is the same as one of the recently introduced schemes of Ref.~\cite{liu_arxiv_frustrated_networks_2023} where the floppy regime and spring-pruning effects were studied. It also bears similarity to the approach taken in Ref.~\cite{massimo_scipost_2023}, however there the interplay between coordination and mechanical frustration was not considered. Finally, our work also echoes the approach of Ref.~\cite{eric_boson_peak_emt} in which mechanical frustration is incorporated into a mean-field, effective-medium theory for glass elasticity, however relaxation towards mechanical equilibrium --- and the consequences of the systems' self-organization during said relaxation --- are not fully captured by that approach.

In Refs.~\cite{scipost_mean_field_qles_2021,meanfield_qle_pierfrancesco_prb_2021} a mean-field model of coupled anharmonic oscillators was formulated and analyzed; that model's vibrational spectrum reproduces two features also seen in the nonphononic spectra of computer glasses: (i) it follows ${\cal D}(\omega)\!=\!A_{\rm g}\omega^4$ in many regions of the model's parameter space, and (ii) the prefactor $A_{\rm g}$ is exponentially suppressed as a function of the relevant combination of the model's parameters, akin to the Boltzmann-like reduction seen in $A_{\rm g}$ of deeply supercooled computer glasses~\cite{pinching_pnas,mw_thermal_origin_of_qle_pre2020} --- with the (equilibrium) parent temperature $T_{\rm p}$ from which glassy samples were (instantaneously) quenched playing the role of the equilibrium temperature in the Boltzmann-like relation. In future work we plan to investigate how the prefactor $A_{\rm g}(z,w)$ behaves as a function of the coordination $z$ and the mechanical frustration $w$ in our minimal model -- to the aim of constructing a mapping between the parameters of the aforementioned mean-field model, and the degree of frustration or proximity to unjamming in finite-dimensional computer glass models. Such a mapping might help shed light on the origin of the universal quartic nonphononic spectra seen in many computer glass models.

Finally, we note that the characteristic mechanical-frustration scale $w_\star\!\sim\!\delta z$ (cf.~discussion in Sect.~\ref{sec:displacements}) is reminiscent of the scaling $\gamma_\star\!\sim\!|\delta z|$ of the stiffening strain $\gamma_\star$ at which floppy, disordered spring networks --- featuring $z\!<\!z_{\rm c}$ --- acquire a finite shear modulus upon subjection to spatial deformations~\cite{mw_maha_prl_2008,gustavo_pre_2014,robbie_pre_2018,strain_stiffening_2023}. Indeed, and as shown in Ref.~\cite{liu_arxiv_frustrated_networks_2023}, a floppy network with a vanishing shear modulus $G\!=\!0$ can be made rigid and acquire a finite shear modulus by introducing mechanical frustration -- in the form of noise in the network's springs' rest-lengths, as done here. Similarly, in Ref.~\cite{chase_motors_prl_2012} it was shown how motor-generated stresses can rigidify floppy elastic networks.

\begin{figure}[ht!]
 \includegraphics[width = 0.5\textwidth]{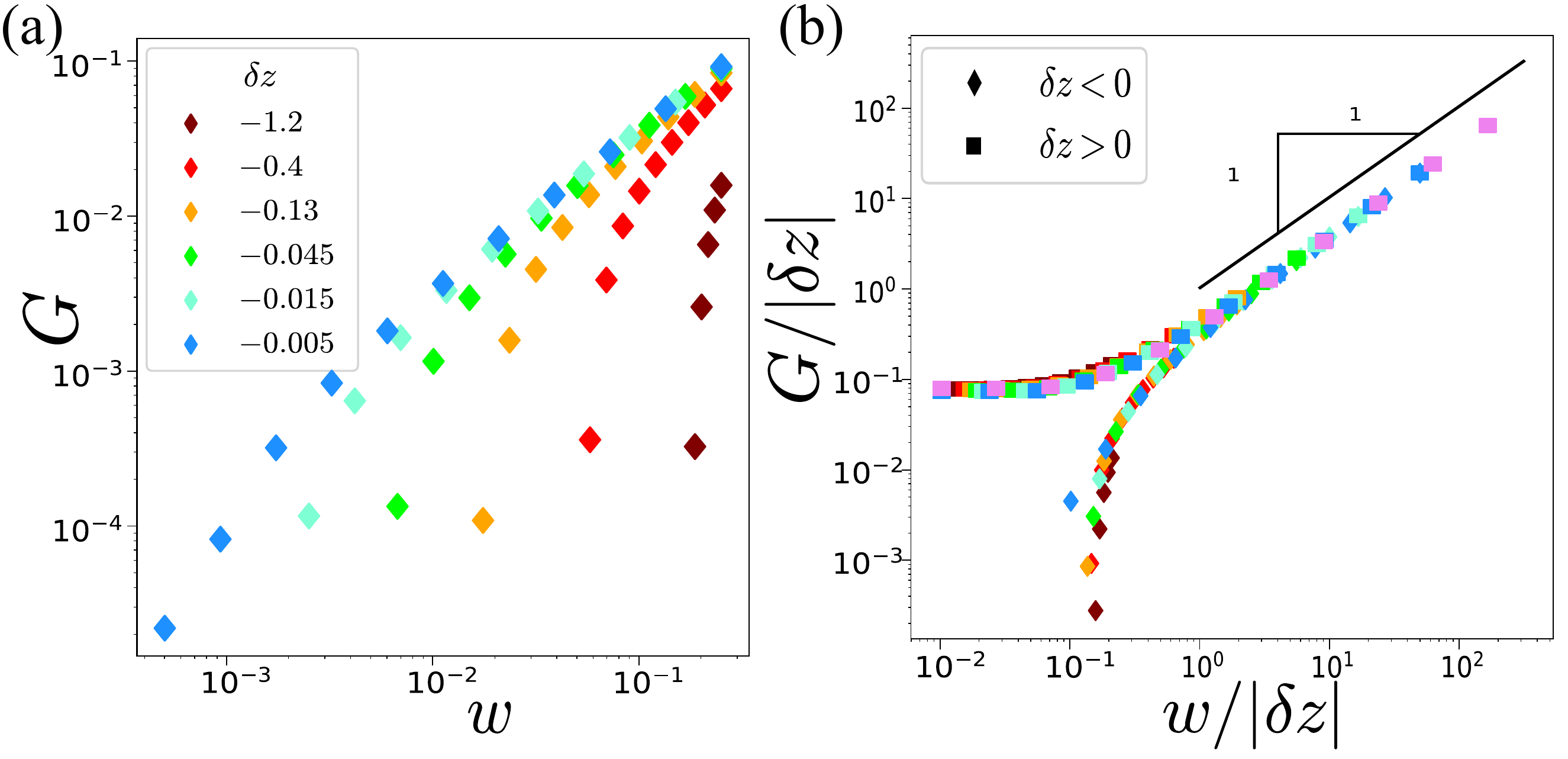}
  \caption{\footnotesize (a) The shear modulus $G$ of hypostatic networks with $z\!<\!z_{\rm c}\!\equiv\!2\dbar$ and $N\!=\!16000$, shown here for various degrees of mechanical frustration $w$. (b) The same scaling form Eq.~(\ref{eq:shear_modulus_scaling_form}) for the shear modulus of hyperstatic networks with $z\!>\!z_{\rm c}$ -- also collapses the shear modulus of frustrated hypostatic networks, and reveals the characteristic frustration scale $w_\star\!\sim\!|\delta z|$ above which $G$ becomes finite, reminiscent of the characteristic stiffening strain scale $\gamma_\star$ above which deformed floppy networks acquire a finite shear modulus, see text for further discussion.}
  \label{fig:floppy_G}
\end{figure}

We demonstrate this effect in Fig.~\ref{fig:floppy_G}, where the shear modulus of floppy networks with $z\!<\!z_{\rm c}$ is plotted vs.~$w$ for a variety of coordinations as detailed by the figure legend. Different from strain-stiffening, in this case our frustrated floppy networks with $w\!<\!|\delta z|$ \emph{can} carry a finite (though small) elastic energy $U/N\!>\!0$, but --- at the same time --- also have a vanishing shear modulus $G\!=\!0$, while strained floppy networks below the stiffening strain --- i.e.~with $\gamma\!<\!\gamma_\star\!\sim\!|\delta z|$ --- have a vanishing shear modulus $G\!=\!0$ but also carry \emph{no} elastic energy, i.e.~$U/N\!=\!0$.   Interestingly, the same scaling variable $w/|\delta z|$ --- put forward and motivated for systems above the jamming point with $z\!>\!z_{\rm c}$ --- collapses the $z\!<\!z_{\rm c}$ data as well, as shown in Fig.~\ref{fig:floppy_G}b, in which we also included the hyperstatic ($z\!>\!z_{\rm c}$) data set. In future work we plan to investigate how the vibrational density of states of  hypostatic samples that were rigidified (i.e.~acquire a finite shear modulus $G\!>\!0$) by frustration -- behaves as compared to hyperstatic frustrated networks. 


\vspace{-0.5cm}

\acknowledgements
\vspace{-0.2cm}
We thank Eran Bouchbinder and Joshua Dijksman for invaluable discussions. This work was supported by the Institute of Physics of the University of Amsterdam.


\begin{thebibliography}{72}%
\makeatletter
\providecommand \@ifxundefined [1]{%
 \@ifx{#1\undefined}
}%
\providecommand \@ifnum [1]{%
 \ifnum #1\expandafter \@firstoftwo
 \else \expandafter \@secondoftwo
 \fi
}%
\providecommand \@ifx [1]{%
 \ifx #1\expandafter \@firstoftwo
 \else \expandafter \@secondoftwo
 \fi
}%
\providecommand \natexlab [1]{#1}%
\providecommand \enquote  [1]{``#1''}%
\providecommand \bibnamefont  [1]{#1}%
\providecommand \bibfnamefont [1]{#1}%
\providecommand \citenamefont [1]{#1}%
\providecommand \href@noop [0]{\@secondoftwo}%
\providecommand \href [0]{\begingroup \@sanitize@url \@href}%
\providecommand \@href[1]{\@@startlink{#1}\@@href}%
\providecommand \@@href[1]{\endgroup#1\@@endlink}%
\providecommand \@sanitize@url [0]{\catcode `\\12\catcode `\$12\catcode
  `\&12\catcode `\#12\catcode `\^12\catcode `\_12\catcode `\%12\relax}%
\providecommand \@@startlink[1]{}%
\providecommand \@@endlink[0]{}%
\providecommand \url  [0]{\begingroup\@sanitize@url \@url }%
\providecommand \@url [1]{\endgroup\@href {#1}{\urlprefix }}%
\providecommand \urlprefix  [0]{URL }%
\providecommand \Eprint [0]{\href }%
\providecommand \doibase [0]{https://doi.org/}%
\providecommand \selectlanguage [0]{\@gobble}%
\providecommand \bibinfo  [0]{\@secondoftwo}%
\providecommand \bibfield  [0]{\@secondoftwo}%
\providecommand \translation [1]{[#1]}%
\providecommand \BibitemOpen [0]{}%
\providecommand \bibitemStop [0]{}%
\providecommand \bibitemNoStop [0]{.\EOS\space}%
\providecommand \EOS [0]{\spacefactor3000\relax}%
\providecommand \BibitemShut  [1]{\csname bibitem#1\endcsname}%
\let\auto@bib@innerbib\@empty
\bibitem [{\citenamefont {Schirmacher}\ \emph {et~al.}(2007)\citenamefont
  {Schirmacher}, \citenamefont {Ruocco},\ and\ \citenamefont
  {Scopigno}}]{Schirmacher_prl_2007}%
  \BibitemOpen
  \bibfield  {author} {\bibinfo {author} {\bibfnamefont {W.}~\bibnamefont
  {Schirmacher}}, \bibinfo {author} {\bibfnamefont {G.}~\bibnamefont
  {Ruocco}},\ and\ \bibinfo {author} {\bibfnamefont {T.}~\bibnamefont
  {Scopigno}},\ }\bibfield  {title} {\bibinfo {title} {Acoustic attenuation in
  glasses and its relation with the boson peak},\ }\href
  {https://doi.org/10.1103/PhysRevLett.98.025501} {\bibfield  {journal}
  {\bibinfo  {journal} {Phys. Rev. Lett.}\ }\textbf {\bibinfo {volume} {98}},\
  \bibinfo {pages} {025501} (\bibinfo {year} {2007})}\BibitemShut {NoStop}%
\bibitem [{\citenamefont {Marruzzo}\ \emph {et~al.}(2013)\citenamefont
  {Marruzzo}, \citenamefont {Schirmacher}, \citenamefont {Fratalocchi},\ and\
  \citenamefont {Ruocco}}]{Schirmacher_2013_boson_peak}%
  \BibitemOpen
  \bibfield  {author} {\bibinfo {author} {\bibfnamefont {A.}~\bibnamefont
  {Marruzzo}}, \bibinfo {author} {\bibfnamefont {W.}~\bibnamefont
  {Schirmacher}}, \bibinfo {author} {\bibfnamefont {A.}~\bibnamefont
  {Fratalocchi}},\ and\ \bibinfo {author} {\bibfnamefont {G.}~\bibnamefont
  {Ruocco}},\ }\bibfield  {title} {\bibinfo {title} {Heterogeneous shear
  elasticity of glasses: the origin of the boson peak},\ }\href
  {https://doi.org/10.1038/srep01407} {\bibfield  {journal} {\bibinfo
  {journal} {Sci. Rep.}\ }\textbf {\bibinfo {volume} {3}},\ \bibinfo {pages}
  {1407 EP } (\bibinfo {year} {2013})}\BibitemShut {NoStop}%
\bibitem [{\citenamefont {Moriel}\ \emph {et~al.}(2019)\citenamefont {Moriel},
  \citenamefont {Kapteijns}, \citenamefont {Rainone}, \citenamefont {Zylberg},
  \citenamefont {Lerner},\ and\ \citenamefont {Bouchbinder}}]{scattering_jcp}%
  \BibitemOpen
  \bibfield  {author} {\bibinfo {author} {\bibfnamefont {A.}~\bibnamefont
  {Moriel}}, \bibinfo {author} {\bibfnamefont {G.}~\bibnamefont {Kapteijns}},
  \bibinfo {author} {\bibfnamefont {C.}~\bibnamefont {Rainone}}, \bibinfo
  {author} {\bibfnamefont {J.}~\bibnamefont {Zylberg}}, \bibinfo {author}
  {\bibfnamefont {E.}~\bibnamefont {Lerner}},\ and\ \bibinfo {author}
  {\bibfnamefont {E.}~\bibnamefont {Bouchbinder}},\ }\bibfield  {title}
  {\bibinfo {title} {Wave attenuation in glasses: Rayleigh and
  generalized-rayleigh scattering scaling},\ }\href
  {https://doi.org/10.1063/1.5111192} {\bibfield  {journal} {\bibinfo
  {journal} {J. Chem. Phys.}\ }\textbf {\bibinfo {volume} {151}},\ \bibinfo
  {pages} {104503} (\bibinfo {year} {2019})}\BibitemShut {NoStop}%
\bibitem [{\citenamefont {Wang}\ \emph {et~al.}(2020)\citenamefont {Wang},
  \citenamefont {Szamel},\ and\ \citenamefont
  {Flenner}}]{grzegorz_soft_matter_2020}%
  \BibitemOpen
  \bibfield  {author} {\bibinfo {author} {\bibfnamefont {L.}~\bibnamefont
  {Wang}}, \bibinfo {author} {\bibfnamefont {G.}~\bibnamefont {Szamel}},\ and\
  \bibinfo {author} {\bibfnamefont {E.}~\bibnamefont {Flenner}},\ }\bibfield
  {title} {\bibinfo {title} {Sound attenuation in finite-temperature stable
  glasses},\ }\href {https://doi.org/10.1039/D0SM00633E} {\bibfield  {journal}
  {\bibinfo  {journal} {Soft Matter}\ }\textbf {\bibinfo {volume} {16}},\
  \bibinfo {pages} {7165} (\bibinfo {year} {2020})}\BibitemShut {NoStop}%
\bibitem [{\citenamefont {Kapteijns}\ \emph
  {et~al.}(2021{\natexlab{a}})\citenamefont {Kapteijns}, \citenamefont
  {Richard}, \citenamefont {Bouchbinder},\ and\ \citenamefont
  {Lerner}}]{jcp_letter_scattering_2021}%
  \BibitemOpen
  \bibfield  {author} {\bibinfo {author} {\bibfnamefont {G.}~\bibnamefont
  {Kapteijns}}, \bibinfo {author} {\bibfnamefont {D.}~\bibnamefont {Richard}},
  \bibinfo {author} {\bibfnamefont {E.}~\bibnamefont {Bouchbinder}},\ and\
  \bibinfo {author} {\bibfnamefont {E.}~\bibnamefont {Lerner}},\ }\bibfield
  {title} {\bibinfo {title} {Elastic moduli fluctuations predict wave
  attenuation rates in glasses},\ }\href {https://doi.org/10.1063/5.0038710}
  {\bibfield  {journal} {\bibinfo  {journal} {J. Chem. Phys.}\ }\textbf
  {\bibinfo {volume} {154}},\ \bibinfo {pages} {081101} (\bibinfo {year}
  {2021}{\natexlab{a}})}\BibitemShut {NoStop}%
\bibitem [{\citenamefont {Mahajan}\ and\ \citenamefont
  {Ciamarra}(2023)}]{massimo_scipost_2023}%
  \BibitemOpen
  \bibfield  {author} {\bibinfo {author} {\bibfnamefont {S.}~\bibnamefont
  {Mahajan}}\ and\ \bibinfo {author} {\bibfnamefont {M.~P.}\ \bibnamefont
  {Ciamarra}},\ }\bibfield  {title} {\bibinfo {title} {{Quasi-localized
  vibrational modes, boson peak and sound attenuation in model mass-spring
  networks}},\ }\href {https://doi.org/10.21468/SciPostPhys.15.2.069}
  {\bibfield  {journal} {\bibinfo  {journal} {SciPost Phys.}\ }\textbf
  {\bibinfo {volume} {15}},\ \bibinfo {pages} {069} (\bibinfo {year}
  {2023})}\BibitemShut {NoStop}%
\bibitem [{\citenamefont {Shi}\ and\ \citenamefont
  {Falk}(2005)}]{falk_shi_prl_2005}%
  \BibitemOpen
  \bibfield  {author} {\bibinfo {author} {\bibfnamefont {Y.}~\bibnamefont
  {Shi}}\ and\ \bibinfo {author} {\bibfnamefont {M.~L.}\ \bibnamefont {Falk}},\
  }\bibfield  {title} {\bibinfo {title} {Strain localization and percolation of
  stable structure in amorphous solids},\ }\href
  {https://doi.org/10.1103/PhysRevLett.95.095502} {\bibfield  {journal}
  {\bibinfo  {journal} {Phys. Rev. Lett.}\ }\textbf {\bibinfo {volume} {95}},\
  \bibinfo {pages} {095502} (\bibinfo {year} {2005})}\BibitemShut {NoStop}%
\bibitem [{\citenamefont {Bouchbinder}\ and\ \citenamefont
  {Langer}(2009{\natexlab{a}})}]{Bouchbinder2009c}%
  \BibitemOpen
  \bibfield  {author} {\bibinfo {author} {\bibfnamefont {E.}~\bibnamefont
  {Bouchbinder}}\ and\ \bibinfo {author} {\bibfnamefont {J.}~\bibnamefont
  {Langer}},\ }\bibfield  {title} {\bibinfo {title} {Nonequilibrium
  thermodynamics of driven amorphous materials. iii. shear-transformation-zone
  plasticity},\ }\href {https://doi.org/10.1103/PhysRevE.80.031133} {\bibfield
  {journal} {\bibinfo  {journal} {Phys. Rev. E}\ }\textbf {\bibinfo {volume}
  {80}},\ \bibinfo {pages} {031133} (\bibinfo {year}
  {2009}{\natexlab{a}})}\BibitemShut {NoStop}%
\bibitem [{\citenamefont {Ozawa}\ \emph {et~al.}(2018)\citenamefont {Ozawa},
  \citenamefont {Berthier}, \citenamefont {Biroli}, \citenamefont {Rosso},\
  and\ \citenamefont {Tarjus}}]{Ozawa6656}%
  \BibitemOpen
  \bibfield  {author} {\bibinfo {author} {\bibfnamefont {M.}~\bibnamefont
  {Ozawa}}, \bibinfo {author} {\bibfnamefont {L.}~\bibnamefont {Berthier}},
  \bibinfo {author} {\bibfnamefont {G.}~\bibnamefont {Biroli}}, \bibinfo
  {author} {\bibfnamefont {A.}~\bibnamefont {Rosso}},\ and\ \bibinfo {author}
  {\bibfnamefont {G.}~\bibnamefont {Tarjus}},\ }\bibfield  {title} {\bibinfo
  {title} {Random critical point separates brittle and ductile yielding
  transitions in amorphous materials},\ }\href
  {https://doi.org/10.1073/pnas.1806156115} {\bibfield  {journal} {\bibinfo
  {journal} {Proc. Natl. Acad. Sci. U.S.A.}\ }\textbf {\bibinfo {volume}
  {115}},\ \bibinfo {pages} {6656} (\bibinfo {year} {2018})}\BibitemShut
  {NoStop}%
\bibitem [{\citenamefont {Jin}\ \emph {et~al.}(2018)\citenamefont {Jin},
  \citenamefont {Urbani}, \citenamefont {Zamponi},\ and\ \citenamefont
  {Yoshino}}]{francescos_and_yoshino_science_advances_2018}%
  \BibitemOpen
  \bibfield  {author} {\bibinfo {author} {\bibfnamefont {Y.}~\bibnamefont
  {Jin}}, \bibinfo {author} {\bibfnamefont {P.}~\bibnamefont {Urbani}},
  \bibinfo {author} {\bibfnamefont {F.}~\bibnamefont {Zamponi}},\ and\ \bibinfo
  {author} {\bibfnamefont {H.}~\bibnamefont {Yoshino}},\ }\bibfield  {title}
  {\bibinfo {title} {A stability-reversibility map unifies elasticity,
  plasticity, yielding, and jamming in hard sphere glasses},\ }\href
  {https://doi.org/10.1126/sciadv.aat6387} {\bibfield  {journal} {\bibinfo
  {journal} {Sci. Adv.}\ }\textbf {\bibinfo {volume} {4}},\ \bibinfo {pages}
  {eaat6387} (\bibinfo {year} {2018})}\BibitemShut {NoStop}%
\bibitem [{\citenamefont {Rycroft}\ and\ \citenamefont
  {Bouchbinder}(2012)}]{Rycroft2012}%
  \BibitemOpen
  \bibfield  {author} {\bibinfo {author} {\bibfnamefont {C.~H.}\ \bibnamefont
  {Rycroft}}\ and\ \bibinfo {author} {\bibfnamefont {E.}~\bibnamefont
  {Bouchbinder}},\ }\bibfield  {title} {\bibinfo {title} {Fracture toughness of
  metallic glasses: Annealing-induced embrittlement},\ }\href
  {https://doi.org/10.1103/PhysRevLett.109.194301} {\bibfield  {journal}
  {\bibinfo  {journal} {Phys. Rev. Lett.}\ }\textbf {\bibinfo {volume} {109}},\
  \bibinfo {pages} {194301} (\bibinfo {year} {2012})}\BibitemShut {NoStop}%
\bibitem [{\citenamefont {Ketkaew}\ \emph {et~al.}(2018)\citenamefont
  {Ketkaew}, \citenamefont {Chen}, \citenamefont {Wang}, \citenamefont {Datye},
  \citenamefont {Fan}, \citenamefont {Pereira}, \citenamefont {Schwarz},
  \citenamefont {Liu}, \citenamefont {Yamada}, \citenamefont {Dmowski},
  \citenamefont {Shattuck}, \citenamefont {O'Hern}, \citenamefont {Egami},
  \citenamefont {Bouchbinder},\ and\ \citenamefont
  {Schroers}}]{Eran_mechanical_glass_transition}%
  \BibitemOpen
  \bibfield  {author} {\bibinfo {author} {\bibfnamefont {J.}~\bibnamefont
  {Ketkaew}}, \bibinfo {author} {\bibfnamefont {W.}~\bibnamefont {Chen}},
  \bibinfo {author} {\bibfnamefont {H.}~\bibnamefont {Wang}}, \bibinfo {author}
  {\bibfnamefont {A.}~\bibnamefont {Datye}}, \bibinfo {author} {\bibfnamefont
  {M.}~\bibnamefont {Fan}}, \bibinfo {author} {\bibfnamefont {G.}~\bibnamefont
  {Pereira}}, \bibinfo {author} {\bibfnamefont {U.~D.}\ \bibnamefont
  {Schwarz}}, \bibinfo {author} {\bibfnamefont {Z.}~\bibnamefont {Liu}},
  \bibinfo {author} {\bibfnamefont {R.}~\bibnamefont {Yamada}}, \bibinfo
  {author} {\bibfnamefont {W.}~\bibnamefont {Dmowski}}, \bibinfo {author}
  {\bibfnamefont {M.~D.}\ \bibnamefont {Shattuck}}, \bibinfo {author}
  {\bibfnamefont {C.~S.}\ \bibnamefont {O'Hern}}, \bibinfo {author}
  {\bibfnamefont {T.}~\bibnamefont {Egami}}, \bibinfo {author} {\bibfnamefont
  {E.}~\bibnamefont {Bouchbinder}},\ and\ \bibinfo {author} {\bibfnamefont
  {J.}~\bibnamefont {Schroers}},\ }\bibfield  {title} {\bibinfo {title}
  {Mechanical glass transition revealed by the fracture toughness of metallic
  glasses},\ }\href {https://doi.org/10.1038/s41467-018-05682-8} {\bibfield
  {journal} {\bibinfo  {journal} {Nat. Commun.}\ }\textbf {\bibinfo {volume}
  {9}},\ \bibinfo {pages} {3271} (\bibinfo {year} {2018})}\BibitemShut
  {NoStop}%
\bibitem [{\citenamefont {Richard}\ \emph {et~al.}(2021)\citenamefont
  {Richard}, \citenamefont {Lerner},\ and\ \citenamefont
  {Bouchbinder}}]{david_fracture_mrs_2021}%
  \BibitemOpen
  \bibfield  {author} {\bibinfo {author} {\bibfnamefont {D.}~\bibnamefont
  {Richard}}, \bibinfo {author} {\bibfnamefont {E.}~\bibnamefont {Lerner}},\
  and\ \bibinfo {author} {\bibfnamefont {E.}~\bibnamefont {Bouchbinder}},\
  }\bibfield  {title} {\bibinfo {title} {Brittle-to-ductile transitions in
  glasses: Roles of soft defects and loading geometry},\ }\href
  {https://doi.org/10.1557/s43577-021-00171-8} {\bibfield  {journal} {\bibinfo
  {journal} {MRS Bulletin}\ }\textbf {\bibinfo {volume} {46}},\ \bibinfo
  {pages} {902} (\bibinfo {year} {2021})}\BibitemShut {NoStop}%
\bibitem [{\citenamefont {Patinet}\ \emph {et~al.}(2016)\citenamefont
  {Patinet}, \citenamefont {Vandembroucq},\ and\ \citenamefont
  {Falk}}]{falk_prl_2016}%
  \BibitemOpen
  \bibfield  {author} {\bibinfo {author} {\bibfnamefont {S.}~\bibnamefont
  {Patinet}}, \bibinfo {author} {\bibfnamefont {D.}~\bibnamefont
  {Vandembroucq}},\ and\ \bibinfo {author} {\bibfnamefont {M.~L.}\ \bibnamefont
  {Falk}},\ }\bibfield  {title} {\bibinfo {title} {Connecting local yield
  stresses with plastic activity in amorphous solids},\ }\href
  {https://doi.org/10.1103/PhysRevLett.117.045501} {\bibfield  {journal}
  {\bibinfo  {journal} {Phys. Rev. Lett.}\ }\textbf {\bibinfo {volume} {117}},\
  \bibinfo {pages} {045501} (\bibinfo {year} {2016})}\BibitemShut {NoStop}%
\bibitem [{\citenamefont {Richard}\ \emph
  {et~al.}(2020{\natexlab{a}})\citenamefont {Richard}, \citenamefont {Ozawa},
  \citenamefont {Patinet}, \citenamefont {Stanifer}, \citenamefont {Shang},
  \citenamefont {Ridout}, \citenamefont {Xu}, \citenamefont {Zhang},
  \citenamefont {Morse}, \citenamefont {Barrat}, \citenamefont {Berthier},
  \citenamefont {Falk}, \citenamefont {Guan}, \citenamefont {Liu},
  \citenamefont {Martens}, \citenamefont {Sastry}, \citenamefont
  {Vandembroucq}, \citenamefont {Lerner},\ and\ \citenamefont
  {Manning}}]{david_collaboration_2020}%
  \BibitemOpen
  \bibfield  {author} {\bibinfo {author} {\bibfnamefont {D.}~\bibnamefont
  {Richard}}, \bibinfo {author} {\bibfnamefont {M.}~\bibnamefont {Ozawa}},
  \bibinfo {author} {\bibfnamefont {S.}~\bibnamefont {Patinet}}, \bibinfo
  {author} {\bibfnamefont {E.}~\bibnamefont {Stanifer}}, \bibinfo {author}
  {\bibfnamefont {B.}~\bibnamefont {Shang}}, \bibinfo {author} {\bibfnamefont
  {S.~A.}\ \bibnamefont {Ridout}}, \bibinfo {author} {\bibfnamefont
  {B.}~\bibnamefont {Xu}}, \bibinfo {author} {\bibfnamefont {G.}~\bibnamefont
  {Zhang}}, \bibinfo {author} {\bibfnamefont {P.~K.}\ \bibnamefont {Morse}},
  \bibinfo {author} {\bibfnamefont {J.-L.}\ \bibnamefont {Barrat}}, \bibinfo
  {author} {\bibfnamefont {L.}~\bibnamefont {Berthier}}, \bibinfo {author}
  {\bibfnamefont {M.~L.}\ \bibnamefont {Falk}}, \bibinfo {author}
  {\bibfnamefont {P.}~\bibnamefont {Guan}}, \bibinfo {author} {\bibfnamefont
  {A.~J.}\ \bibnamefont {Liu}}, \bibinfo {author} {\bibfnamefont
  {K.}~\bibnamefont {Martens}}, \bibinfo {author} {\bibfnamefont
  {S.}~\bibnamefont {Sastry}}, \bibinfo {author} {\bibfnamefont
  {D.}~\bibnamefont {Vandembroucq}}, \bibinfo {author} {\bibfnamefont
  {E.}~\bibnamefont {Lerner}},\ and\ \bibinfo {author} {\bibfnamefont {M.~L.}\
  \bibnamefont {Manning}},\ }\bibfield  {title} {\bibinfo {title} {Predicting
  plasticity in disordered solids from structural indicators},\ }\href
  {https://doi.org/10.1103/PhysRevMaterials.4.113609} {\bibfield  {journal}
  {\bibinfo  {journal} {Phys. Rev. Materials}\ }\textbf {\bibinfo {volume}
  {4}},\ \bibinfo {pages} {113609} (\bibinfo {year}
  {2020}{\natexlab{a}})}\BibitemShut {NoStop}%
\bibitem [{\citenamefont {Gonz\'alez-L\'opez}\ \emph
  {et~al.}(2023)\citenamefont {Gonz\'alez-L\'opez}, \citenamefont
  {Bouchbinder},\ and\ \citenamefont {Lerner}}]{chi_paper_2023}%
  \BibitemOpen
  \bibfield  {author} {\bibinfo {author} {\bibfnamefont {K.}~\bibnamefont
  {Gonz\'alez-L\'opez}}, \bibinfo {author} {\bibfnamefont {E.}~\bibnamefont
  {Bouchbinder}},\ and\ \bibinfo {author} {\bibfnamefont {E.}~\bibnamefont
  {Lerner}},\ }\bibfield  {title} {\bibinfo {title} {Variability of mesoscopic
  mechanical disorder in disordered solids},\ }\href
  {https://doi.org/https://doi.org/10.1016/j.jnoncrysol.2023.122137} {\bibfield
   {journal} {\bibinfo  {journal} {J. Non-Cryst. Solids.}\ }\textbf {\bibinfo
  {volume} {604}},\ \bibinfo {pages} {122137} (\bibinfo {year}
  {2023})}\BibitemShut {NoStop}%
\bibitem [{\citenamefont {Jung}\ \emph {et~al.}(2023)\citenamefont {Jung},
  \citenamefont {Alkemade}, \citenamefont {Bapst}, \citenamefont {Coslovich},
  \citenamefont {Filion}, \citenamefont {Landes}, \citenamefont {Liu},
  \citenamefont {Pezzicoli}, \citenamefont {Shiba}, \citenamefont {Volpe} \emph
  {et~al.}}]{ml_roadmap_2023}%
  \BibitemOpen
  \bibfield  {author} {\bibinfo {author} {\bibfnamefont {G.}~\bibnamefont
  {Jung}}, \bibinfo {author} {\bibfnamefont {R.~M.}\ \bibnamefont {Alkemade}},
  \bibinfo {author} {\bibfnamefont {V.}~\bibnamefont {Bapst}}, \bibinfo
  {author} {\bibfnamefont {D.}~\bibnamefont {Coslovich}}, \bibinfo {author}
  {\bibfnamefont {L.}~\bibnamefont {Filion}}, \bibinfo {author} {\bibfnamefont
  {F.~P.}\ \bibnamefont {Landes}}, \bibinfo {author} {\bibfnamefont
  {A.}~\bibnamefont {Liu}}, \bibinfo {author} {\bibfnamefont {F.~S.}\
  \bibnamefont {Pezzicoli}}, \bibinfo {author} {\bibfnamefont {H.}~\bibnamefont
  {Shiba}}, \bibinfo {author} {\bibfnamefont {G.}~\bibnamefont {Volpe}}, \emph
  {et~al.},\ }\bibfield  {title} {\bibinfo {title} {Roadmap on machine learning
  glassy liquids},\ }\href {https://arxiv.org/abs/2311.14752} {\bibfield
  {journal} {\bibinfo  {journal} {arXiv preprint arXiv:2311.14752}\ } (\bibinfo
  {year} {2023})}\BibitemShut {NoStop}%
\bibitem [{\citenamefont {Richard}(2023)}]{david_pairwise_stuff_2023}%
  \BibitemOpen
  \bibfield  {author} {\bibinfo {author} {\bibfnamefont {D.}~\bibnamefont
  {Richard}},\ }\bibfield  {title} {\bibinfo {title} {Connecting microscopic
  and mesoscopic mechanics in model structural glasses},\ }\href
  {https://arxiv.org/abs/2311.09917} {\bibfield  {journal} {\bibinfo  {journal}
  {arXiv preprint arXiv:2311.09917}\ } (\bibinfo {year} {2023})}\BibitemShut
  {NoStop}%
\bibitem [{\citenamefont {Royall}\ and\ \citenamefont
  {Williams}(2015)}]{paddy_huge_review_2015}%
  \BibitemOpen
  \bibfield  {author} {\bibinfo {author} {\bibfnamefont {C.~P.}\ \bibnamefont
  {Royall}}\ and\ \bibinfo {author} {\bibfnamefont {S.~R.}\ \bibnamefont
  {Williams}},\ }\bibfield  {title} {\bibinfo {title} {The role of local
  structure in dynamical arrest},\ }\href
  {https://doi.org/https://doi.org/10.1016/j.physrep.2014.11.004} {\bibfield
  {journal} {\bibinfo  {journal} {Phys. Rep.}\ }\textbf {\bibinfo {volume}
  {560}},\ \bibinfo {pages} {1 } (\bibinfo {year} {2015})}\BibitemShut
  {NoStop}%
\bibitem [{\citenamefont {Tong}\ and\ \citenamefont {Tanaka}(2019)}]{Tong2019}%
  \BibitemOpen
  \bibfield  {author} {\bibinfo {author} {\bibfnamefont {H.}~\bibnamefont
  {Tong}}\ and\ \bibinfo {author} {\bibfnamefont {H.}~\bibnamefont {Tanaka}},\
  }\bibfield  {title} {\bibinfo {title} {Structural order as a genuine control
  parameter of dynamics in simple glass formers},\ }\href
  {https://doi.org/10.1038/s41467-019-13606-3} {\bibfield  {journal} {\bibinfo
  {journal} {Nature Communications}\ }\textbf {\bibinfo {volume} {10}},\
  \bibinfo {pages} {5596} (\bibinfo {year} {2019})}\BibitemShut {NoStop}%
\bibitem [{\citenamefont {Bouchbinder}\ and\ \citenamefont
  {Langer}(2009{\natexlab{b}})}]{Bouchbinder2009b}%
  \BibitemOpen
  \bibfield  {author} {\bibinfo {author} {\bibfnamefont {E.}~\bibnamefont
  {Bouchbinder}}\ and\ \bibinfo {author} {\bibfnamefont {J.}~\bibnamefont
  {Langer}},\ }\bibfield  {title} {\bibinfo {title} {Nonequilibrium
  thermodynamics of driven amorphous materials. ii. effective-temperature
  theory},\ }\href {https://doi.org/10.1103/PhysRevE.80.031132} {\bibfield
  {journal} {\bibinfo  {journal} {Phys. Rev. E}\ }\textbf {\bibinfo {volume}
  {80}},\ \bibinfo {pages} {031132} (\bibinfo {year}
  {2009}{\natexlab{b}})}\BibitemShut {NoStop}%
\bibitem [{\citenamefont {Brito}\ \emph {et~al.}(2018)\citenamefont {Brito},
  \citenamefont {Lerner},\ and\ \citenamefont {Wyart}}]{swap_prx_MW}%
  \BibitemOpen
  \bibfield  {author} {\bibinfo {author} {\bibfnamefont {C.}~\bibnamefont
  {Brito}}, \bibinfo {author} {\bibfnamefont {E.}~\bibnamefont {Lerner}},\ and\
  \bibinfo {author} {\bibfnamefont {M.}~\bibnamefont {Wyart}},\ }\bibfield
  {title} {\bibinfo {title} {Theory for swap acceleration near the glass and
  jamming transitions for continuously polydisperse particles},\ }\href
  {https://doi.org/10.1103/PhysRevX.8.031050} {\bibfield  {journal} {\bibinfo
  {journal} {Phys. Rev. X}\ }\textbf {\bibinfo {volume} {8}},\ \bibinfo {pages}
  {031050} (\bibinfo {year} {2018})}\BibitemShut {NoStop}%
\bibitem [{\citenamefont {Kapteijns}\ \emph {et~al.}(2019)\citenamefont
  {Kapteijns}, \citenamefont {Ji}, \citenamefont {Brito}, \citenamefont
  {Wyart},\ and\ \citenamefont {Lerner}}]{fsp}%
  \BibitemOpen
  \bibfield  {author} {\bibinfo {author} {\bibfnamefont {G.}~\bibnamefont
  {Kapteijns}}, \bibinfo {author} {\bibfnamefont {W.}~\bibnamefont {Ji}},
  \bibinfo {author} {\bibfnamefont {C.}~\bibnamefont {Brito}}, \bibinfo
  {author} {\bibfnamefont {M.}~\bibnamefont {Wyart}},\ and\ \bibinfo {author}
  {\bibfnamefont {E.}~\bibnamefont {Lerner}},\ }\bibfield  {title} {\bibinfo
  {title} {Fast generation of ultrastable computer glasses by minimization of
  an augmented potential energy},\ }\href
  {https://doi.org/10.1103/PhysRevE.99.012106} {\bibfield  {journal} {\bibinfo
  {journal} {Phys. Rev. E}\ }\textbf {\bibinfo {volume} {99}},\ \bibinfo
  {pages} {012106} (\bibinfo {year} {2019})}\BibitemShut {NoStop}%
\bibitem [{\citenamefont {Lerner}\ and\ \citenamefont
  {Bouchbinder}(2018{\natexlab{a}})}]{inst_note}%
  \BibitemOpen
  \bibfield  {author} {\bibinfo {author} {\bibfnamefont {E.}~\bibnamefont
  {Lerner}}\ and\ \bibinfo {author} {\bibfnamefont {E.}~\bibnamefont
  {Bouchbinder}},\ }\bibfield  {title} {\bibinfo {title} {Frustration-induced
  internal stresses are responsible for quasilocalized modes in structural
  glasses},\ }\href {https://doi.org/10.1103/PhysRevE.97.032140} {\bibfield
  {journal} {\bibinfo  {journal} {Phys. Rev. E}\ }\textbf {\bibinfo {volume}
  {97}},\ \bibinfo {pages} {032140} (\bibinfo {year}
  {2018}{\natexlab{a}})}\BibitemShut {NoStop}%
\bibitem [{\citenamefont {Ninarello}\ \emph {et~al.}(2017)\citenamefont
  {Ninarello}, \citenamefont {Berthier},\ and\ \citenamefont
  {Coslovich}}]{LB_swap_prx}%
  \BibitemOpen
  \bibfield  {author} {\bibinfo {author} {\bibfnamefont {A.}~\bibnamefont
  {Ninarello}}, \bibinfo {author} {\bibfnamefont {L.}~\bibnamefont
  {Berthier}},\ and\ \bibinfo {author} {\bibfnamefont {D.}~\bibnamefont
  {Coslovich}},\ }\bibfield  {title} {\bibinfo {title} {Models and algorithms
  for the next generation of glass transition studies},\ }\href
  {https://doi.org/10.1103/PhysRevX.7.021039} {\bibfield  {journal} {\bibinfo
  {journal} {Phys. Rev. X}\ }\textbf {\bibinfo {volume} {7}},\ \bibinfo {pages}
  {021039} (\bibinfo {year} {2017})}\BibitemShut {NoStop}%
\bibitem [{\citenamefont {Kapteijns}\ \emph
  {et~al.}(2021{\natexlab{b}})\citenamefont {Kapteijns}, \citenamefont
  {Bouchbinder},\ and\ \citenamefont {Lerner}}]{phonon_widths2}%
  \BibitemOpen
  \bibfield  {author} {\bibinfo {author} {\bibfnamefont {G.}~\bibnamefont
  {Kapteijns}}, \bibinfo {author} {\bibfnamefont {E.}~\bibnamefont
  {Bouchbinder}},\ and\ \bibinfo {author} {\bibfnamefont {E.}~\bibnamefont
  {Lerner}},\ }\bibfield  {title} {\bibinfo {title} {Unified quantifier of
  mechanical disorder in solids},\ }\href
  {https://doi.org/10.1103/PhysRevE.104.035001} {\bibfield  {journal} {\bibinfo
   {journal} {Phys. Rev. E}\ }\textbf {\bibinfo {volume} {104}},\ \bibinfo
  {pages} {035001} (\bibinfo {year} {2021}{\natexlab{b}})}\BibitemShut
  {NoStop}%
\bibitem [{\citenamefont {DeGiuli}\ \emph {et~al.}(2014)\citenamefont
  {DeGiuli}, \citenamefont {Laversanne-Finot}, \citenamefont {During},
  \citenamefont {Lerner},\ and\ \citenamefont {Wyart}}]{eric_boson_peak_emt}%
  \BibitemOpen
  \bibfield  {author} {\bibinfo {author} {\bibfnamefont {E.}~\bibnamefont
  {DeGiuli}}, \bibinfo {author} {\bibfnamefont {A.}~\bibnamefont
  {Laversanne-Finot}}, \bibinfo {author} {\bibfnamefont {G.}~\bibnamefont
  {During}}, \bibinfo {author} {\bibfnamefont {E.}~\bibnamefont {Lerner}},\
  and\ \bibinfo {author} {\bibfnamefont {M.}~\bibnamefont {Wyart}},\ }\bibfield
   {title} {\bibinfo {title} {Effects of coordination and pressure on sound
  attenuation{,} boson peak and elasticity in amorphous solids},\ }\href
  {https://doi.org/10.1039/C4SM00561A} {\bibfield  {journal} {\bibinfo
  {journal} {Soft Matter}\ }\textbf {\bibinfo {volume} {10}},\ \bibinfo {pages}
  {5628} (\bibinfo {year} {2014})}\BibitemShut {NoStop}%
\bibitem [{\citenamefont {Lerner}\ \emph {et~al.}(2014)\citenamefont {Lerner},
  \citenamefont {DeGiuli}, \citenamefont {During},\ and\ \citenamefont
  {Wyart}}]{breakdown}%
  \BibitemOpen
  \bibfield  {author} {\bibinfo {author} {\bibfnamefont {E.}~\bibnamefont
  {Lerner}}, \bibinfo {author} {\bibfnamefont {E.}~\bibnamefont {DeGiuli}},
  \bibinfo {author} {\bibfnamefont {G.}~\bibnamefont {During}},\ and\ \bibinfo
  {author} {\bibfnamefont {M.}~\bibnamefont {Wyart}},\ }\bibfield  {title}
  {\bibinfo {title} {Breakdown of continuum elasticity in amorphous solids},\
  }\href {https://doi.org/10.1039/C4SM00311J} {\bibfield  {journal} {\bibinfo
  {journal} {Soft Matter}\ }\textbf {\bibinfo {volume} {10}},\ \bibinfo {pages}
  {5085} (\bibinfo {year} {2014})}\BibitemShut {NoStop}%
\bibitem [{\citenamefont {Karpov}\ and\ \citenamefont
  {Parshin}(1985)}]{soft_potential_model_01}%
  \BibitemOpen
  \bibfield  {author} {\bibinfo {author} {\bibfnamefont {V.}~\bibnamefont
  {Karpov}}\ and\ \bibinfo {author} {\bibfnamefont {D.}~\bibnamefont
  {Parshin}},\ }\bibfield  {title} {\bibinfo {title} {On the thermal
  conductivity of glasses at temperatures below the debye temperature},\ }\href
  {http://jetp.ras.ru/cgi-bin/dn/e_061_06_1308.pdf} {\bibfield  {journal}
  {\bibinfo  {journal} {Zh. Eksp. Teor. Fiz}\ }\textbf {\bibinfo {volume}
  {88}},\ \bibinfo {pages} {2212} (\bibinfo {year} {1985})}\BibitemShut
  {NoStop}%
\bibitem [{\citenamefont {Ilyin}\ \emph {et~al.}(1987)\citenamefont {Ilyin},
  \citenamefont {Karpov},\ and\ \citenamefont
  {Parshin}}]{soft_potential_model_02}%
  \BibitemOpen
  \bibfield  {author} {\bibinfo {author} {\bibfnamefont {M.}~\bibnamefont
  {Ilyin}}, \bibinfo {author} {\bibfnamefont {V.}~\bibnamefont {Karpov}},\ and\
  \bibinfo {author} {\bibfnamefont {D.}~\bibnamefont {Parshin}},\ }\bibfield
  {title} {\bibinfo {title} {Parameters of soft atomic potentials in glasses},\
  }\href {http://jetp.ras.ru/cgi-bin/dn/e_065_01_0165.pdf} {\bibfield
  {journal} {\bibinfo  {journal} {Zh. Eksp. Teor. Fiz.}\ }\textbf {\bibinfo
  {volume} {92}},\ \bibinfo {pages} {291} (\bibinfo {year} {1987})}\BibitemShut
  {NoStop}%
\bibitem [{\citenamefont {Lerner}\ and\ \citenamefont
  {Bouchbinder}(2021)}]{JCP_Perspective}%
  \BibitemOpen
  \bibfield  {author} {\bibinfo {author} {\bibfnamefont {E.}~\bibnamefont
  {Lerner}}\ and\ \bibinfo {author} {\bibfnamefont {E.}~\bibnamefont
  {Bouchbinder}},\ }\bibfield  {title} {\bibinfo {title} {Low-energy
  quasilocalized excitations in structural glasses},\ }\href
  {https://doi.org/10.1063/5.0069477} {\bibfield  {journal} {\bibinfo
  {journal} {J. Chem. Phys.}\ }\textbf {\bibinfo {volume} {155}},\ \bibinfo
  {pages} {200901} (\bibinfo {year} {2021})}\BibitemShut {NoStop}%
\bibitem [{\citenamefont {Wang}\ \emph {et~al.}(2019)\citenamefont {Wang},
  \citenamefont {Ninarello}, \citenamefont {Guan}, \citenamefont {Berthier},
  \citenamefont {Szamel},\ and\ \citenamefont {Flenner}}]{LB_modes_2019}%
  \BibitemOpen
  \bibfield  {author} {\bibinfo {author} {\bibfnamefont {L.}~\bibnamefont
  {Wang}}, \bibinfo {author} {\bibfnamefont {A.}~\bibnamefont {Ninarello}},
  \bibinfo {author} {\bibfnamefont {P.}~\bibnamefont {Guan}}, \bibinfo {author}
  {\bibfnamefont {L.}~\bibnamefont {Berthier}}, \bibinfo {author}
  {\bibfnamefont {G.}~\bibnamefont {Szamel}},\ and\ \bibinfo {author}
  {\bibfnamefont {E.}~\bibnamefont {Flenner}},\ }\bibfield  {title} {\bibinfo
  {title} {Low-frequency vibrational modes of stable glasses},\ }\href
  {https://doi.org/10.1038/s41467-018-07978-1} {\bibfield  {journal} {\bibinfo
  {journal} {Nat. Commun.}\ }\textbf {\bibinfo {volume} {10}},\ \bibinfo
  {pages} {26} (\bibinfo {year} {2019})}\BibitemShut {NoStop}%
\bibitem [{\citenamefont {Rainone}\ \emph {et~al.}(2020)\citenamefont
  {Rainone}, \citenamefont {Bouchbinder},\ and\ \citenamefont
  {Lerner}}]{pinching_pnas}%
  \BibitemOpen
  \bibfield  {author} {\bibinfo {author} {\bibfnamefont {C.}~\bibnamefont
  {Rainone}}, \bibinfo {author} {\bibfnamefont {E.}~\bibnamefont
  {Bouchbinder}},\ and\ \bibinfo {author} {\bibfnamefont {E.}~\bibnamefont
  {Lerner}},\ }\bibfield  {title} {\bibinfo {title} {Pinching a glass reveals
  key properties of its soft spots},\ }\href
  {https://doi.org/10.1073/pnas.1919958117} {\bibfield  {journal} {\bibinfo
  {journal} {Proc. Natl. Acad. Sci. U.S.A.}\ }\textbf {\bibinfo {volume}
  {117}},\ \bibinfo {pages} {5228} (\bibinfo {year} {2020})}\BibitemShut
  {NoStop}%
\bibitem [{foo()}]{footnote_finite_size_effects}%
  \BibitemOpen
  \bibinfo {note} {Discussions about glass-formation-protocol-induced
  finite-size effects in the nonphononic spectrum of computer glasses can be
  found in
  Refs.~\cite{lerner2019finite,grzegorz_erratum_2022,2d_spectra_jcp_2022,wang_nonsense_jcp_2022,grzegorz_2d_modes_jcp_2023}.}\BibitemShut
  {Stop}%
\bibitem [{\citenamefont {Kapteijns}\ \emph {et~al.}(2018)\citenamefont
  {Kapteijns}, \citenamefont {Bouchbinder},\ and\ \citenamefont
  {Lerner}}]{modes_prl_2018}%
  \BibitemOpen
  \bibfield  {author} {\bibinfo {author} {\bibfnamefont {G.}~\bibnamefont
  {Kapteijns}}, \bibinfo {author} {\bibfnamefont {E.}~\bibnamefont
  {Bouchbinder}},\ and\ \bibinfo {author} {\bibfnamefont {E.}~\bibnamefont
  {Lerner}},\ }\bibfield  {title} {\bibinfo {title} {Universal nonphononic
  density of states in 2d, 3d, and 4d glasses},\ }\href
  {https://doi.org/10.1103/PhysRevLett.121.055501} {\bibfield  {journal}
  {\bibinfo  {journal} {Phys. Rev. Lett.}\ }\textbf {\bibinfo {volume} {121}},\
  \bibinfo {pages} {055501} (\bibinfo {year} {2018})}\BibitemShut {NoStop}%
\bibitem [{\citenamefont {Richard}\ \emph
  {et~al.}(2020{\natexlab{b}})\citenamefont {Richard}, \citenamefont
  {Gonz\'alez-L\'opez}, \citenamefont {Kapteijns}, \citenamefont {Pater},
  \citenamefont {Vaknin}, \citenamefont {Bouchbinder},\ and\ \citenamefont
  {Lerner}}]{modes_prl_2020}%
  \BibitemOpen
  \bibfield  {author} {\bibinfo {author} {\bibfnamefont {D.}~\bibnamefont
  {Richard}}, \bibinfo {author} {\bibfnamefont {K.}~\bibnamefont
  {Gonz\'alez-L\'opez}}, \bibinfo {author} {\bibfnamefont {G.}~\bibnamefont
  {Kapteijns}}, \bibinfo {author} {\bibfnamefont {R.}~\bibnamefont {Pater}},
  \bibinfo {author} {\bibfnamefont {T.}~\bibnamefont {Vaknin}}, \bibinfo
  {author} {\bibfnamefont {E.}~\bibnamefont {Bouchbinder}},\ and\ \bibinfo
  {author} {\bibfnamefont {E.}~\bibnamefont {Lerner}},\ }\bibfield  {title}
  {\bibinfo {title} {Universality of the nonphononic vibrational spectrum
  across different classes of computer glasses},\ }\href
  {https://doi.org/10.1103/PhysRevLett.125.085502} {\bibfield  {journal}
  {\bibinfo  {journal} {Phys. Rev. Lett.}\ }\textbf {\bibinfo {volume} {125}},\
  \bibinfo {pages} {085502} (\bibinfo {year} {2020}{\natexlab{b}})}\BibitemShut
  {NoStop}%
\bibitem [{\citenamefont {Rainone}\ \emph {et~al.}(2021)\citenamefont
  {Rainone}, \citenamefont {Urbani}, \citenamefont {Zamponi}, \citenamefont
  {Lerner},\ and\ \citenamefont {Bouchbinder}}]{scipost_mean_field_qles_2021}%
  \BibitemOpen
  \bibfield  {author} {\bibinfo {author} {\bibfnamefont {C.}~\bibnamefont
  {Rainone}}, \bibinfo {author} {\bibfnamefont {P.}~\bibnamefont {Urbani}},
  \bibinfo {author} {\bibfnamefont {F.}~\bibnamefont {Zamponi}}, \bibinfo
  {author} {\bibfnamefont {E.}~\bibnamefont {Lerner}},\ and\ \bibinfo {author}
  {\bibfnamefont {E.}~\bibnamefont {Bouchbinder}},\ }\bibfield  {title}
  {\bibinfo {title} {{Mean-field model of interacting quasilocalized
  excitations in glasses}},\ }\href
  {https://doi.org/10.21468/SciPostPhysCore.4.2.008} {\bibfield  {journal}
  {\bibinfo  {journal} {SciPost Phys. Core}\ }\textbf {\bibinfo {volume} {4}},\
  \bibinfo {pages} {8} (\bibinfo {year} {2021})}\BibitemShut {NoStop}%
\bibitem [{\citenamefont {Bouchbinder}\ \emph {et~al.}(2021)\citenamefont
  {Bouchbinder}, \citenamefont {Lerner}, \citenamefont {Rainone}, \citenamefont
  {Urbani},\ and\ \citenamefont
  {Zamponi}}]{meanfield_qle_pierfrancesco_prb_2021}%
  \BibitemOpen
  \bibfield  {author} {\bibinfo {author} {\bibfnamefont {E.}~\bibnamefont
  {Bouchbinder}}, \bibinfo {author} {\bibfnamefont {E.}~\bibnamefont {Lerner}},
  \bibinfo {author} {\bibfnamefont {C.}~\bibnamefont {Rainone}}, \bibinfo
  {author} {\bibfnamefont {P.}~\bibnamefont {Urbani}},\ and\ \bibinfo {author}
  {\bibfnamefont {F.}~\bibnamefont {Zamponi}},\ }\bibfield  {title} {\bibinfo
  {title} {Low-frequency vibrational spectrum of mean-field disordered
  systems},\ }\href {https://doi.org/10.1103/PhysRevB.103.174202} {\bibfield
  {journal} {\bibinfo  {journal} {Phys. Rev. B}\ }\textbf {\bibinfo {volume}
  {103}},\ \bibinfo {pages} {174202} (\bibinfo {year} {2021})}\BibitemShut
  {NoStop}%
\bibitem [{\citenamefont {Sharma}\ \emph {et~al.}(2016)\citenamefont {Sharma},
  \citenamefont {Licup}, \citenamefont {Jansen}, \citenamefont {Rens},
  \citenamefont {Sheinman}, \citenamefont {Koenderink},\ and\ \citenamefont
  {MacKintosh}}]{robbie_nature_physics_2016}%
  \BibitemOpen
  \bibfield  {author} {\bibinfo {author} {\bibfnamefont {A.}~\bibnamefont
  {Sharma}}, \bibinfo {author} {\bibfnamefont {A.~J.}\ \bibnamefont {Licup}},
  \bibinfo {author} {\bibfnamefont {K.~A.}\ \bibnamefont {Jansen}}, \bibinfo
  {author} {\bibfnamefont {R.}~\bibnamefont {Rens}}, \bibinfo {author}
  {\bibfnamefont {M.}~\bibnamefont {Sheinman}}, \bibinfo {author}
  {\bibfnamefont {G.~H.}\ \bibnamefont {Koenderink}},\ and\ \bibinfo {author}
  {\bibfnamefont {F.~C.}\ \bibnamefont {MacKintosh}},\ }\bibfield  {title}
  {\bibinfo {title} {Strain-controlled criticality governs the nonlinear
  mechanics of fibre networks},\ }\href {https://doi.org/10.1038/nphys3628}
  {\bibfield  {journal} {\bibinfo  {journal} {Nature Physics}\ }\textbf
  {\bibinfo {volume} {12}},\ \bibinfo {pages} {584} (\bibinfo {year}
  {2016})}\BibitemShut {NoStop}%
\bibitem [{\citenamefont {Rens}\ \emph {et~al.}(2018)\citenamefont {Rens},
  \citenamefont {Villarroel}, \citenamefont {D\"uring},\ and\ \citenamefont
  {Lerner}}]{robbie_pre_2018}%
  \BibitemOpen
  \bibfield  {author} {\bibinfo {author} {\bibfnamefont {R.}~\bibnamefont
  {Rens}}, \bibinfo {author} {\bibfnamefont {C.}~\bibnamefont {Villarroel}},
  \bibinfo {author} {\bibfnamefont {G.}~\bibnamefont {D\"uring}},\ and\
  \bibinfo {author} {\bibfnamefont {E.}~\bibnamefont {Lerner}},\ }\bibfield
  {title} {\bibinfo {title} {Micromechanical theory of strain stiffening of
  biopolymer networks},\ }\href {https://doi.org/10.1103/PhysRevE.98.062411}
  {\bibfield  {journal} {\bibinfo  {journal} {Phys. Rev. E}\ }\textbf {\bibinfo
  {volume} {98}},\ \bibinfo {pages} {062411} (\bibinfo {year}
  {2018})}\BibitemShut {NoStop}%
\bibitem [{\citenamefont {Lerner}\ and\ \citenamefont
  {Bouchbinder}(2023{\natexlab{a}})}]{strain_stiffening_2023}%
  \BibitemOpen
  \bibfield  {author} {\bibinfo {author} {\bibfnamefont {E.}~\bibnamefont
  {Lerner}}\ and\ \bibinfo {author} {\bibfnamefont {E.}~\bibnamefont
  {Bouchbinder}},\ }\bibfield  {title} {\bibinfo {title} {Scaling theory of
  critical strain-stiffening in disordered elastic networks},\ }\href
  {https://doi.org/https://doi.org/10.1016/j.eml.2023.102104} {\bibfield
  {journal} {\bibinfo  {journal} {Extreme Mechanics Letters}\ }\textbf
  {\bibinfo {volume} {65}},\ \bibinfo {pages} {102104} (\bibinfo {year}
  {2023}{\natexlab{a}})}\BibitemShut {NoStop}%
\bibitem [{\citenamefont {Lerner}\ and\ \citenamefont
  {Bouchbinder}(2023{\natexlab{b}})}]{anomalous_elasticity_soft_matter_2023}%
  \BibitemOpen
  \bibfield  {author} {\bibinfo {author} {\bibfnamefont {E.}~\bibnamefont
  {Lerner}}\ and\ \bibinfo {author} {\bibfnamefont {E.}~\bibnamefont
  {Bouchbinder}},\ }\bibfield  {title} {\bibinfo {title} {Anomalous linear
  elasticity of disordered networks},\ }\href
  {https://doi.org/10.1039/D2SM01253G} {\bibfield  {journal} {\bibinfo
  {journal} {Soft Matter}\ }\textbf {\bibinfo {volume} {19}},\ \bibinfo {pages}
  {1076} (\bibinfo {year} {2023}{\natexlab{b}})}\BibitemShut {NoStop}%
\bibitem [{\citenamefont {Bitzek}\ \emph {et~al.}(2006)\citenamefont {Bitzek},
  \citenamefont {Koskinen}, \citenamefont {G\"ahler}, \citenamefont {Moseler},\
  and\ \citenamefont {Gumbsch}}]{fire}%
  \BibitemOpen
  \bibfield  {author} {\bibinfo {author} {\bibfnamefont {E.}~\bibnamefont
  {Bitzek}}, \bibinfo {author} {\bibfnamefont {P.}~\bibnamefont {Koskinen}},
  \bibinfo {author} {\bibfnamefont {F.}~\bibnamefont {G\"ahler}}, \bibinfo
  {author} {\bibfnamefont {M.}~\bibnamefont {Moseler}},\ and\ \bibinfo {author}
  {\bibfnamefont {P.}~\bibnamefont {Gumbsch}},\ }\bibfield  {title} {\bibinfo
  {title} {Structural relaxation made simple},\ }\href
  {https://doi.org/10.1103/PhysRevLett.97.170201} {\bibfield  {journal}
  {\bibinfo  {journal} {Phys. Rev. Lett.}\ }\textbf {\bibinfo {volume} {97}},\
  \bibinfo {pages} {170201} (\bibinfo {year} {2006})}\BibitemShut {NoStop}%
\bibitem [{\citenamefont {Cunha}\ \emph {et~al.}(2023)\citenamefont {Cunha},
  \citenamefont {Crocker},\ and\ \citenamefont
  {Liu}}]{liu_arxiv_frustrated_networks_2023}%
  \BibitemOpen
  \bibfield  {author} {\bibinfo {author} {\bibfnamefont {M.~A.~G.}\
  \bibnamefont {Cunha}}, \bibinfo {author} {\bibfnamefont {J.~C.}\ \bibnamefont
  {Crocker}},\ and\ \bibinfo {author} {\bibfnamefont {A.~J.}\ \bibnamefont
  {Liu}},\ }\bibfield  {title} {\bibinfo {title} {Building rigid networks with
  prestress and selective pruning},\ }\href {https://arxiv.org/abs/2312.06119}
  {\bibfield  {journal} {\bibinfo  {journal} {arXiv preprint arXiv:2312.06119}\
  } (\bibinfo {year} {2023})}\BibitemShut {NoStop}%
\bibitem [{\citenamefont {Lerner}\ \emph {et~al.}(2012)\citenamefont {Lerner},
  \citenamefont {D\"uring},\ and\ \citenamefont {Wyart}}]{asm_pnas_2012}%
  \BibitemOpen
  \bibfield  {author} {\bibinfo {author} {\bibfnamefont {E.}~\bibnamefont
  {Lerner}}, \bibinfo {author} {\bibfnamefont {G.}~\bibnamefont {D\"uring}},\
  and\ \bibinfo {author} {\bibfnamefont {M.}~\bibnamefont {Wyart}},\ }\bibfield
   {title} {\bibinfo {title} {A unified framework for non-brownian suspension
  flows and soft amorphous solids},\ }\href
  {https://doi.org/10.1073/pnas.1120215109} {\bibfield  {journal} {\bibinfo
  {journal} {Proc. Natl. Acad. Sci. U.S.A.}\ }\textbf {\bibinfo {volume}
  {109}},\ \bibinfo {pages} {4798} (\bibinfo {year} {2012})}\BibitemShut
  {NoStop}%
\bibitem [{\citenamefont {Lerner}(2018)}]{sss_epje_2018}%
  \BibitemOpen
  \bibfield  {author} {\bibinfo {author} {\bibfnamefont {E.}~\bibnamefont
  {Lerner}},\ }\bibfield  {title} {\bibinfo {title} {Quasilocalized states of
  self stress in packing-derived networks},\ }\href
  {https://doi.org/10.1140/epje/i2018-11705-9} {\bibfield  {journal} {\bibinfo
  {journal} {Eur. Phys. J. E}\ }\textbf {\bibinfo {volume} {41}},\ \bibinfo
  {pages} {93} (\bibinfo {year} {2018})}\BibitemShut {NoStop}%
\bibitem [{\citenamefont {Calladine}(1978)}]{calladine_buckminster_1978}%
  \BibitemOpen
  \bibfield  {author} {\bibinfo {author} {\bibfnamefont {C.~R.}\ \bibnamefont
  {Calladine}},\ }\bibfield  {title} {\bibinfo {title} {Buckminster {Fuller}'s
  “{Tensegrity}” structures and {Clerk} {Maxwell}'s rules for the
  construction of stiff frames},\ }\href
  {https://doi.org/10.1016/0020-7683(78)90052-5} {\bibfield  {journal}
  {\bibinfo  {journal} {International Journal of Solids and Structures}\
  }\textbf {\bibinfo {volume} {14}},\ \bibinfo {pages} {161} (\bibinfo {year}
  {1978})}\BibitemShut {NoStop}%
\bibitem [{\citenamefont {Wyart}(2005)}]{matthieu_thesis}%
  \BibitemOpen
  \bibfield  {author} {\bibinfo {author} {\bibfnamefont {M.}~\bibnamefont
  {Wyart}},\ }\bibfield  {title} {\bibinfo {title} {On the rigidity of
  amorphous solids},\ }\href {https://doi.org/10.1051/anphys:2006003}
  {\bibfield  {journal} {\bibinfo  {journal} {Ann. Phys. Fr.}\ }\textbf
  {\bibinfo {volume} {30}},\ \bibinfo {pages} {1} (\bibinfo {year}
  {2005})}\BibitemShut {NoStop}%
\bibitem [{\citenamefont {D{\"u}ring}\ \emph {et~al.}(2013)\citenamefont
  {D{\"u}ring}, \citenamefont {Lerner},\ and\ \citenamefont
  {Wyart}}]{phonon_gap_2012}%
  \BibitemOpen
  \bibfield  {author} {\bibinfo {author} {\bibfnamefont {G.}~\bibnamefont
  {D{\"u}ring}}, \bibinfo {author} {\bibfnamefont {E.}~\bibnamefont {Lerner}},\
  and\ \bibinfo {author} {\bibfnamefont {M.}~\bibnamefont {Wyart}},\ }\bibfield
   {title} {\bibinfo {title} {Phonon gap and localization lengths in floppy
  materials},\ }\href {https://doi.org/10.1039/C2SM25878A} {\bibfield
  {journal} {\bibinfo  {journal} {Soft Matter}\ }\textbf {\bibinfo {volume}
  {9}},\ \bibinfo {pages} {146} (\bibinfo {year} {2013})}\BibitemShut {NoStop}%
\bibitem [{\citenamefont {Beltukov}(2015)}]{Beltukov2015}%
  \BibitemOpen
  \bibfield  {author} {\bibinfo {author} {\bibfnamefont {Y.~M.}\ \bibnamefont
  {Beltukov}},\ }\bibfield  {title} {\bibinfo {title} {Random matrix theory
  approach to vibrations near the jamming transition},\ }\href
  {https://doi.org/10.1134/S0021364015050033} {\bibfield  {journal} {\bibinfo
  {journal} {JETP Letters}\ }\textbf {\bibinfo {volume} {101}},\ \bibinfo
  {pages} {345} (\bibinfo {year} {2015})}\BibitemShut {NoStop}%
\bibitem [{\citenamefont {Mizuno}\ \emph {et~al.}(2016)\citenamefont {Mizuno},
  \citenamefont {Saitoh},\ and\ \citenamefont {Silbert}}]{silbert_pre_2016}%
  \BibitemOpen
  \bibfield  {author} {\bibinfo {author} {\bibfnamefont {H.}~\bibnamefont
  {Mizuno}}, \bibinfo {author} {\bibfnamefont {K.}~\bibnamefont {Saitoh}},\
  and\ \bibinfo {author} {\bibfnamefont {L.~E.}\ \bibnamefont {Silbert}},\
  }\bibfield  {title} {\bibinfo {title} {Elastic moduli and vibrational modes
  in jammed particulate packings},\ }\href
  {https://doi.org/10.1103/PhysRevE.93.062905} {\bibfield  {journal} {\bibinfo
  {journal} {Phys. Rev. E}\ }\textbf {\bibinfo {volume} {93}},\ \bibinfo
  {pages} {062905} (\bibinfo {year} {2016})}\BibitemShut {NoStop}%
\bibitem [{\citenamefont {Lutsko}(1989)}]{lutsko}%
  \BibitemOpen
  \bibfield  {author} {\bibinfo {author} {\bibfnamefont {J.~F.}\ \bibnamefont
  {Lutsko}},\ }\bibfield  {title} {\bibinfo {title} {Generalized expressions
  for the calculation of elastic constants by computer simulation},\ }\href
  {https://doi.org/10.1063/1.342716} {\bibfield  {journal} {\bibinfo  {journal}
  {J. Appl. Phys.}\ }\textbf {\bibinfo {volume} {65}},\ \bibinfo {pages} {2991}
  (\bibinfo {year} {1989})}\BibitemShut {NoStop}%
\bibitem [{\citenamefont {Ellenbroek}\ \emph {et~al.}(2009)\citenamefont
  {Ellenbroek}, \citenamefont {Zeravcic}, \citenamefont {van Saarloos},\ and\
  \citenamefont {van Hecke}}]{Ellenbroek_2009}%
  \BibitemOpen
  \bibfield  {author} {\bibinfo {author} {\bibfnamefont {W.~G.}\ \bibnamefont
  {Ellenbroek}}, \bibinfo {author} {\bibfnamefont {Z.}~\bibnamefont
  {Zeravcic}}, \bibinfo {author} {\bibfnamefont {W.}~\bibnamefont {van
  Saarloos}},\ and\ \bibinfo {author} {\bibfnamefont {M.}~\bibnamefont {van
  Hecke}},\ }\bibfield  {title} {\bibinfo {title} {Non-affine response: Jammed
  packings vs. spring networks},\ }\href
  {https://doi.org/10.1209/0295-5075/87/34004} {\bibfield  {journal} {\bibinfo
  {journal} {Europhys. Lett.}\ }\textbf {\bibinfo {volume} {87}},\ \bibinfo
  {pages} {34004} (\bibinfo {year} {2009})}\BibitemShut {NoStop}%
\bibitem [{\citenamefont {Lerner}\ and\ \citenamefont
  {Bouchbinder}(2018{\natexlab{b}})}]{cge_paper}%
  \BibitemOpen
  \bibfield  {author} {\bibinfo {author} {\bibfnamefont {E.}~\bibnamefont
  {Lerner}}\ and\ \bibinfo {author} {\bibfnamefont {E.}~\bibnamefont
  {Bouchbinder}},\ }\bibfield  {title} {\bibinfo {title} {A characteristic
  energy scale in glasses},\ }\href {https://doi.org/10.1063/1.5024776}
  {\bibfield  {journal} {\bibinfo  {journal} {J. Chem. Phys.}\ }\textbf
  {\bibinfo {volume} {148}},\ \bibinfo {pages} {214502} (\bibinfo {year}
  {2018}{\natexlab{b}})}\BibitemShut {NoStop}%
\bibitem [{\citenamefont {Gonz\'alez-L\'opez}\ \emph
  {et~al.}(2021)\citenamefont {Gonz\'alez-L\'opez}, \citenamefont {Shivam},
  \citenamefont {Zheng}, \citenamefont {Ciamarra},\ and\ \citenamefont
  {Lerner}}]{sticky_spheres_part2}%
  \BibitemOpen
  \bibfield  {author} {\bibinfo {author} {\bibfnamefont {K.}~\bibnamefont
  {Gonz\'alez-L\'opez}}, \bibinfo {author} {\bibfnamefont {M.}~\bibnamefont
  {Shivam}}, \bibinfo {author} {\bibfnamefont {Y.}~\bibnamefont {Zheng}},
  \bibinfo {author} {\bibfnamefont {M.~P.}\ \bibnamefont {Ciamarra}},\ and\
  \bibinfo {author} {\bibfnamefont {E.}~\bibnamefont {Lerner}},\ }\bibfield
  {title} {\bibinfo {title} {Mechanical disorder of sticky-sphere glasses.
  {II}. thermomechanical inannealability},\ }\href
  {https://doi.org/10.1103/PhysRevE.103.022606} {\bibfield  {journal} {\bibinfo
   {journal} {Phys. Rev. E}\ }\textbf {\bibinfo {volume} {103}},\ \bibinfo
  {pages} {022606} (\bibinfo {year} {2021})}\BibitemShut {NoStop}%
\bibitem [{\citenamefont
  {Wyart}(2010{\natexlab{a}})}]{wyart_vibrational_entropy}%
  \BibitemOpen
  \bibfield  {author} {\bibinfo {author} {\bibfnamefont {M.}~\bibnamefont
  {Wyart}},\ }\bibfield  {title} {\bibinfo {title} {Correlations between
  vibrational entropy and dynamics in liquids},\ }\href
  {https://doi.org/10.1103/PhysRevLett.104.095901} {\bibfield  {journal}
  {\bibinfo  {journal} {Phys. Rev. Lett.}\ }\textbf {\bibinfo {volume} {104}},\
  \bibinfo {pages} {095901} (\bibinfo {year} {2010}{\natexlab{a}})}\BibitemShut
  {NoStop}%
\bibitem [{\citenamefont {Wyart}\ \emph {et~al.}(2005)\citenamefont {Wyart},
  \citenamefont {Silbert}, \citenamefont {Nagel},\ and\ \citenamefont
  {Witten}}]{matthieu_PRE_2005}%
  \BibitemOpen
  \bibfield  {author} {\bibinfo {author} {\bibfnamefont {M.}~\bibnamefont
  {Wyart}}, \bibinfo {author} {\bibfnamefont {L.~E.}\ \bibnamefont {Silbert}},
  \bibinfo {author} {\bibfnamefont {S.~R.}\ \bibnamefont {Nagel}},\ and\
  \bibinfo {author} {\bibfnamefont {T.~A.}\ \bibnamefont {Witten}},\ }\bibfield
   {title} {\bibinfo {title} {Effects of compression on the vibrational modes
  of marginally jammed solids},\ }\href
  {https://doi.org/10.1103/PhysRevE.72.051306} {\bibfield  {journal} {\bibinfo
  {journal} {Phys. Rev. E}\ }\textbf {\bibinfo {volume} {72}},\ \bibinfo
  {pages} {051306} (\bibinfo {year} {2005})}\BibitemShut {NoStop}%
\bibitem [{\citenamefont {Wyart}(2010{\natexlab{b}})}]{mw_EMT_epl}%
  \BibitemOpen
  \bibfield  {author} {\bibinfo {author} {\bibfnamefont {M.}~\bibnamefont
  {Wyart}},\ }\bibfield  {title} {\bibinfo {title} {Scaling of phononic
  transport with connectivity in amorphous solids},\ }\href
  {http://stacks.iop.org/0295-5075/89/i=6/a=64001} {\bibfield  {journal}
  {\bibinfo  {journal} {Europhys. Lett.}\ }\textbf {\bibinfo {volume} {89}},\
  \bibinfo {pages} {64001} (\bibinfo {year} {2010}{\natexlab{b}})}\BibitemShut
  {NoStop}%
\bibitem [{\citenamefont {Yan}\ \emph {et~al.}(2016)\citenamefont {Yan},
  \citenamefont {DeGiuli},\ and\ \citenamefont
  {Wyart}}]{new_variational_argument_epl_2016}%
  \BibitemOpen
  \bibfield  {author} {\bibinfo {author} {\bibfnamefont {L.}~\bibnamefont
  {Yan}}, \bibinfo {author} {\bibfnamefont {E.}~\bibnamefont {DeGiuli}},\ and\
  \bibinfo {author} {\bibfnamefont {M.}~\bibnamefont {Wyart}},\ }\bibfield
  {title} {\bibinfo {title} {On variational arguments for vibrational modes
  near jamming},\ }\href {http://stacks.iop.org/0295-5075/114/i=2/a=26003}
  {\bibfield  {journal} {\bibinfo  {journal} {Europhys. Lett.}\ }\textbf
  {\bibinfo {volume} {114}},\ \bibinfo {pages} {26003} (\bibinfo {year}
  {2016})}\BibitemShut {NoStop}%
\bibitem [{\citenamefont {Lerner}(2023)}]{bending_networks_arXiv_2023}%
  \BibitemOpen
  \bibfield  {author} {\bibinfo {author} {\bibfnamefont {E.}~\bibnamefont
  {Lerner}},\ }\bibfield  {title} {\bibinfo {title} {Effects of coordination
  and stiffness scale-separation in disordered elastic networks},\ }\href
  {https://arxiv.org/abs/2312.06427} {\bibfield  {journal} {\bibinfo  {journal}
  {arXiv preprint arXiv:2312.06427}\ } (\bibinfo {year} {2023})}\BibitemShut
  {NoStop}%
\bibitem [{\citenamefont {Bouchbinder}\ and\ \citenamefont
  {Lerner}(2018)}]{phonon_widths}%
  \BibitemOpen
  \bibfield  {author} {\bibinfo {author} {\bibfnamefont {E.}~\bibnamefont
  {Bouchbinder}}\ and\ \bibinfo {author} {\bibfnamefont {E.}~\bibnamefont
  {Lerner}},\ }\bibfield  {title} {\bibinfo {title} {Universal disorder-induced
  broadening of phonon bands: from disordered lattices to glasses},\ }\href
  {http://stacks.iop.org/1367-2630/20/i=7/a=073022} {\bibfield  {journal}
  {\bibinfo  {journal} {New J. Phys.}\ }\textbf {\bibinfo {volume} {20}},\
  \bibinfo {pages} {073022} (\bibinfo {year} {2018})}\BibitemShut {NoStop}%
\bibitem [{\citenamefont {Giannini}\ \emph {et~al.}(2024)\citenamefont
  {Giannini}, \citenamefont {Lerner}, \citenamefont {Zamponi},\ and\
  \citenamefont {Manning}}]{julia_chi_jcp}%
  \BibitemOpen
  \bibfield  {author} {\bibinfo {author} {\bibfnamefont {J.~A.}\ \bibnamefont
  {Giannini}}, \bibinfo {author} {\bibfnamefont {E.}~\bibnamefont {Lerner}},
  \bibinfo {author} {\bibfnamefont {F.}~\bibnamefont {Zamponi}},\ and\ \bibinfo
  {author} {\bibfnamefont {M.~L.}\ \bibnamefont {Manning}},\ }\bibfield
  {title} {\bibinfo {title} {{Scaling regimes and fluctuations of observables
  in computer glasses approaching the unjamming transition}},\ }\href
  {https://doi.org/10.1063/5.0176713} {\bibfield  {journal} {\bibinfo
  {journal} {J. Chem. Phys.}\ }\textbf {\bibinfo {volume} {160}},\ \bibinfo
  {pages} {034502} (\bibinfo {year} {2024})}\BibitemShut {NoStop}%
\bibitem [{\citenamefont {Kapteijns}\ \emph
  {et~al.}(2021{\natexlab{c}})\citenamefont {Kapteijns}, \citenamefont
  {Bouchbinder},\ and\ \citenamefont {Lerner}}]{phonon_width_2}%
  \BibitemOpen
  \bibfield  {author} {\bibinfo {author} {\bibfnamefont {G.}~\bibnamefont
  {Kapteijns}}, \bibinfo {author} {\bibfnamefont {E.}~\bibnamefont
  {Bouchbinder}},\ and\ \bibinfo {author} {\bibfnamefont {E.}~\bibnamefont
  {Lerner}},\ }\bibfield  {title} {\bibinfo {title} {Unified quantifier of
  mechanical disorder in solids},\ }\href
  {https://doi.org/10.1103/PhysRevE.104.035001} {\bibfield  {journal} {\bibinfo
   {journal} {Phys. Rev. E}\ }\textbf {\bibinfo {volume} {104}},\ \bibinfo
  {pages} {035001} (\bibinfo {year} {2021}{\natexlab{c}})}\BibitemShut
  {NoStop}%
\bibitem [{\citenamefont {Ji}\ \emph {et~al.}(2020)\citenamefont {Ji},
  \citenamefont {de~Geus}, \citenamefont {Popovi\ifmmode~\acute{c}\else
  \'{c}\fi{}}, \citenamefont {Agoritsas},\ and\ \citenamefont
  {Wyart}}]{mw_thermal_origin_of_qle_pre2020}%
  \BibitemOpen
  \bibfield  {author} {\bibinfo {author} {\bibfnamefont {W.}~\bibnamefont
  {Ji}}, \bibinfo {author} {\bibfnamefont {T.~W.~J.}\ \bibnamefont {de~Geus}},
  \bibinfo {author} {\bibfnamefont {M.}~\bibnamefont
  {Popovi\ifmmode~\acute{c}\else \'{c}\fi{}}}, \bibinfo {author} {\bibfnamefont
  {E.}~\bibnamefont {Agoritsas}},\ and\ \bibinfo {author} {\bibfnamefont
  {M.}~\bibnamefont {Wyart}},\ }\bibfield  {title} {\bibinfo {title} {Thermal
  origin of quasilocalized excitations in glasses},\ }\href
  {https://doi.org/10.1103/PhysRevE.102.062110} {\bibfield  {journal} {\bibinfo
   {journal} {Phys. Rev. E}\ }\textbf {\bibinfo {volume} {102}},\ \bibinfo
  {pages} {062110} (\bibinfo {year} {2020})}\BibitemShut {NoStop}%
\bibitem [{\citenamefont {Wyart}\ \emph {et~al.}(2008)\citenamefont {Wyart},
  \citenamefont {Liang}, \citenamefont {Kabla},\ and\ \citenamefont
  {Mahadevan}}]{mw_maha_prl_2008}%
  \BibitemOpen
  \bibfield  {author} {\bibinfo {author} {\bibfnamefont {M.}~\bibnamefont
  {Wyart}}, \bibinfo {author} {\bibfnamefont {H.}~\bibnamefont {Liang}},
  \bibinfo {author} {\bibfnamefont {A.}~\bibnamefont {Kabla}},\ and\ \bibinfo
  {author} {\bibfnamefont {L.}~\bibnamefont {Mahadevan}},\ }\bibfield  {title}
  {\bibinfo {title} {Elasticity of floppy and stiff random networks},\ }\href
  {https://doi.org/10.1103/PhysRevLett.101.215501} {\bibfield  {journal}
  {\bibinfo  {journal} {Phys. Rev. Lett.}\ }\textbf {\bibinfo {volume} {101}},\
  \bibinfo {pages} {215501} (\bibinfo {year} {2008})}\BibitemShut {NoStop}%
\bibitem [{\citenamefont {D\"uring}\ \emph {et~al.}(2014)\citenamefont
  {D\"uring}, \citenamefont {Lerner},\ and\ \citenamefont
  {Wyart}}]{gustavo_pre_2014}%
  \BibitemOpen
  \bibfield  {author} {\bibinfo {author} {\bibfnamefont {G.}~\bibnamefont
  {D\"uring}}, \bibinfo {author} {\bibfnamefont {E.}~\bibnamefont {Lerner}},\
  and\ \bibinfo {author} {\bibfnamefont {M.}~\bibnamefont {Wyart}},\ }\bibfield
   {title} {\bibinfo {title} {Length scales and self-organization in dense
  suspension flows},\ }\href {https://doi.org/10.1103/PhysRevE.89.022305}
  {\bibfield  {journal} {\bibinfo  {journal} {Phys. Rev. E}\ }\textbf {\bibinfo
  {volume} {89}},\ \bibinfo {pages} {022305} (\bibinfo {year}
  {2014})}\BibitemShut {NoStop}%
\bibitem [{\citenamefont {Sheinman}\ \emph {et~al.}(2012)\citenamefont
  {Sheinman}, \citenamefont {Broedersz},\ and\ \citenamefont
  {MacKintosh}}]{chase_motors_prl_2012}%
  \BibitemOpen
  \bibfield  {author} {\bibinfo {author} {\bibfnamefont {M.}~\bibnamefont
  {Sheinman}}, \bibinfo {author} {\bibfnamefont {C.~P.}\ \bibnamefont
  {Broedersz}},\ and\ \bibinfo {author} {\bibfnamefont {F.~C.}\ \bibnamefont
  {MacKintosh}},\ }\bibfield  {title} {\bibinfo {title} {Actively stressed
  marginal networks},\ }\href {https://doi.org/10.1103/PhysRevLett.109.238101}
  {\bibfield  {journal} {\bibinfo  {journal} {Phys. Rev. Lett.}\ }\textbf
  {\bibinfo {volume} {109}},\ \bibinfo {pages} {238101} (\bibinfo {year}
  {2012})}\BibitemShut {NoStop}%
\bibitem [{\citenamefont {Lerner}(2020)}]{lerner2019finite}%
  \BibitemOpen
  \bibfield  {author} {\bibinfo {author} {\bibfnamefont {E.}~\bibnamefont
  {Lerner}},\ }\bibfield  {title} {\bibinfo {title} {Finite-size effects in the
  nonphononic density of states in computer glasses},\ }\href
  {https://doi.org/10.1103/PhysRevE.101.032120} {\bibfield  {journal} {\bibinfo
   {journal} {Phys. Rev. E}\ }\textbf {\bibinfo {volume} {101}},\ \bibinfo
  {pages} {032120} (\bibinfo {year} {2020})}\BibitemShut {NoStop}%
\bibitem [{\citenamefont {Wang}\ \emph
  {et~al.}(2022{\natexlab{a}})\citenamefont {Wang}, \citenamefont {Szamel},\
  and\ \citenamefont {Flenner}}]{grzegorz_erratum_2022}%
  \BibitemOpen
  \bibfield  {author} {\bibinfo {author} {\bibfnamefont {L.}~\bibnamefont
  {Wang}}, \bibinfo {author} {\bibfnamefont {G.}~\bibnamefont {Szamel}},\ and\
  \bibinfo {author} {\bibfnamefont {E.}~\bibnamefont {Flenner}},\ }\bibfield
  {title} {\bibinfo {title} {Erratum: Low-frequency excess vibrational modes in
  two-dimensional glasses [physical review letters 127, 248001 (2021)]},\
  }\href {https://doi.org/10.1103/PhysRevLett.129.019901} {\bibfield  {journal}
  {\bibinfo  {journal} {Phys. Rev. Lett.}\ }\textbf {\bibinfo {volume} {129}},\
  \bibinfo {pages} {019901} (\bibinfo {year} {2022}{\natexlab{a}})}\BibitemShut
  {NoStop}%
\bibitem [{\citenamefont {Lerner}\ and\ \citenamefont
  {Bouchbinder}(2022)}]{2d_spectra_jcp_2022}%
  \BibitemOpen
  \bibfield  {author} {\bibinfo {author} {\bibfnamefont {E.}~\bibnamefont
  {Lerner}}\ and\ \bibinfo {author} {\bibfnamefont {E.}~\bibnamefont
  {Bouchbinder}},\ }\bibfield  {title} {\bibinfo {title} {{Nonphononic spectrum
  of two-dimensional structural glasses}},\ }\href
  {https://doi.org/10.1063/5.0120115} {\bibfield  {journal} {\bibinfo
  {journal} {J. Chem. Phys.}\ }\textbf {\bibinfo {volume} {157}},\ \bibinfo
  {pages} {166101} (\bibinfo {year} {2022})}\BibitemShut {NoStop}%
\bibitem [{\citenamefont {Wang}\ \emph
  {et~al.}(2022{\natexlab{b}})\citenamefont {Wang}, \citenamefont {Fu},\ and\
  \citenamefont {Nie}}]{wang_nonsense_jcp_2022}%
  \BibitemOpen
  \bibfield  {author} {\bibinfo {author} {\bibfnamefont {L.}~\bibnamefont
  {Wang}}, \bibinfo {author} {\bibfnamefont {L.}~\bibnamefont {Fu}},\ and\
  \bibinfo {author} {\bibfnamefont {Y.}~\bibnamefont {Nie}},\ }\bibfield
  {title} {\bibinfo {title} {{Density of states below the first sound mode in
  3D glasses}},\ }\href {https://doi.org/10.1063/5.0102081} {\bibfield
  {journal} {\bibinfo  {journal} {J. Chem. Phys.}\ }\textbf {\bibinfo {volume}
  {157}},\ \bibinfo {pages} {074502} (\bibinfo {year}
  {2022}{\natexlab{b}})}\BibitemShut {NoStop}%
\bibitem [{\citenamefont {Wang}\ \emph {et~al.}(2023)\citenamefont {Wang},
  \citenamefont {Szamel},\ and\ \citenamefont
  {Flenner}}]{grzegorz_2d_modes_jcp_2023}%
  \BibitemOpen
  \bibfield  {author} {\bibinfo {author} {\bibfnamefont {L.}~\bibnamefont
  {Wang}}, \bibinfo {author} {\bibfnamefont {G.}~\bibnamefont {Szamel}},\ and\
  \bibinfo {author} {\bibfnamefont {E.}~\bibnamefont {Flenner}},\ }\bibfield
  {title} {\bibinfo {title} {{Scaling of the non-phononic spectrum of
  two-dimensional glasses}},\ }\href {https://doi.org/10.1063/5.0139596}
  {\bibfield  {journal} {\bibinfo  {journal} {J. Chem. Phys.}\ }\textbf
  {\bibinfo {volume} {158}},\ \bibinfo {pages} {126101} (\bibinfo {year}
  {2023})}\BibitemShut {NoStop}%
\end{thebibliography}
%

\end{document}